\documentclass{aa}
\usepackage{epsf}
\begin{document}

%
   \title{Density waves in the shearing sheet}
   \subtitle{I. Swing amplification}

   \author{B. Fuchs}


   \institute{Astronomisches Rechen--Institut,
              M\"onchhofstra{\ss}e 12--14, 69120 Heidelberg, Germany\\
              e-mail: {\tt fuchs@ari.uni-heidelberg.de}}

   \date{Received 6 November 2000/ Accepted 14 December 2000}

   \abstract{
   The shearing sheet model of a galactic disk is studied anew.
   The theoretical description of its dynamics is based on three
   building blocks: Stellar orbits, which are described here in epicyclic
   approximation, the collisionless Boltzmann equation determining the
   distribution function of stars in phase space, and the Poisson equation in
   order to take account of the self--gravity of the disk. Using these
   tools I develop a new formalism to describe perturbations of the shearing
   sheet. Applying this to the unbounded shearing sheet model I demonstrate
   again how the disturbances of the disk evolve always into `swing
   amplified' density waves, i.e.~spiral--arm like, shearing density
   enhancements, which grow and decay while the wave crests swing by from
   leading to trailing orientation. Several examples are given how such `swing
   amplification' events are incited in the shearing sheet.
   }
\maketitle

%

\section{Introduction}

The shearing sheet (Goldreich \& Lynden--Bell 1965, Julian \& Toomre 1966)
model has been developed as a tool to study the dynamics of galactic disks and
is particularly well suited to describe theoretically the dynamical mechanisms
responsible for the
formation of spiral arms. For the sake of simplicity the model describes only
the dynamics of a patch of a galactic disk. It is assumed to be infinitesimally 
thin and its radial size is assumed to be
much smaller than the disk. Polar coordinates can be
therefore rectified to pseudo-cartesian coordinates and the velocity field of
the differential rotation of the disk can be approximated by a linear shear
flow. These simplifications allow an analytical treatment of the problem,
which helps to clarify the underlying physical processes operating in the disk.
Goldreich \& Lynden--Bell (1965) used hydrodynamical equations as
dynamical equations, which were combined with the Poisson equation in order to
take account of the self-gravity of the disk. Their principal result was a
gravitational instability giving rise to spiral--arm like density enhancements,
which take part in the general shear of the disk and grow while swinging by
from leading to trailing orientation. This effect has been pursued and
discussed further in many studies, notably by Goldreich \& Tremaine (1978).
Independently Julian \& Toomre (1966) have investigated a 
shearing sheet made of stars by using the collisionless Boltzmann equation
combined with the Poisson equation and found also `swing amplification' of
spiral density waves with
the additional effect that the density perturbations die out eventually due to
smearing out of the ever tightening spiral arms by the finite epicyclic
motions of the stars. Toomre (1981) has argued vigorously that this 
swing--amplification mechanism is the principal process of spiral arm formation
in galactic disks and that the spiral arms of many spiral galaxies are thus of 
transient nature. Indeed, numerical simulations of the dynamical evolution of 
self gravitating, differentially rotating disks by Sellwood \& Carlberg (1984)
have shown impressively recurring swing--amplification events, which lead to an
ever--changing appearance of the disk.

The alternative concept of quasi--stationary density waves has been pursued in 
the very influential work of Lin, Shu, Bertin and collaborators culminating 
in the papers by Bertin et al.~(1989 a,b) and Lowe et al.~(1994). In their work
they always consider full disk models. One key ingredient of their recent models
is the introduction of an inner reflecting Q--barrier, which shields the inner 
Lindblad resonance. Together with the semi--reflecting corotation zone this 
forms a kind of resonance cavity, in which the rigidly rotating density waves 
grow exponentially. 

A general theory of modes of star disks has been developed
by Kalnajs in a series of papers (Kalnajs 1971, 1977, 1978), which has been used
by him and several other authors, notably Hunter (1992) and Pichon \& Cannon 
(1997) to describe spiral and bar--like perturbations of the disks. Density 
waves as modes of disks have been also searched for and localized in numerical
simulations of the dynamical evolution of star disks by Donner \& Thomasson
(1994) and Sellwood and collaborators (see Sellwood \& Evans 2000 and 
references therein, also Toomre 1981).
Toomre and Kalnajs (1980) have introduced inner boundaries into their numerical,
softened gravity, simulations of the evolution of the shearing 
sheet and describe two kind of modes growing in the sheet.
The present paper is the first of a series meant as a complement to these 
studies. I consider again a shearing sheet made of stars
and discuss first the unbounded sheet and then the dynamical consequences, when 
inner boundary conditions are applied. The aim is to give a consistent
theoretical description of pure swing--amplification as well as exponentially 
growing modes in the framework of the same model.
In this paper I consider the unbounded shearing sheet. The purpose is to 
develop the 
formalism, and I obtain mainly previously known results again. Contrary to 
previous studies I avoid Lagrangian -- shearing -- coordinates and stay in the
Euler picture. This leads to a compact, transparent form of the central 
Volterra integral equation (cf.~section 6), and will enable me also later, when
I consider a bounded shearing sheet, to derive a dispersion relation for 
rigidly rotating, exponentially growing modes of the disk.

\section{Stellar orbits}

The first step required in order to construct a stellardynamical model of a 
galactic disk is to describe the stellar orbits in the disk. The equations of
motion of the stars are derived from the Lagrangian
\begin{equation}
L = \frac{1}{2} \left( \dot{r}^{\rm{2}} + r \dot{\theta}^{\rm{2}} \right) - 
\Phi\left(r,\theta,z\right)\,,
\end{equation}
where $r$, $\theta$ denote polar coordinates. Since the disk is assumed to be
infinitesimally thin, only plane orbits are considered. $\dot{r}$,
$\dot{\theta}$ are the corresponding radial and angular velocities of the stars.
$\Phi(r,\theta,z)$ is the gravitational potential of the disk. The potential is
assumed to be axisymmetric, $\Phi(r,z)$, in the present section. 
Non--axisymmetric perturbations of this basic state of the disk will be taken 
into account in the next section. Considering now a patch of the disk as in the
shearing sheet model I introduce pseudo--cartesian coordinates
\begin{equation}
x = r - r_{\rm{0}} \quad {\rm and} \quad y = r_{\rm{0}}\left(\theta - 
\Omega_{\rm{0}} t\right)\,.
\end{equation}
$r_0$ denotes the galactocentric distance of the center of the patch and 
$\Omega_0 = \left[\frac{1}{r_0}\left( \frac {d \Phi}{dr} \right)_{\rm r_0} 
\right]^{\frac{1}{2}}$ is the mean angular velocity of the patch around the 
galactic center. In order to keep the analysis tractable I consider the 
stellar orbits in epicyclic approximation (Lindblad 1959). This assumes that
\begin{equation}
|x| \ll r_0 \quad {\rm and} \quad |\dot{x}|,|\dot{y}| \ll \Omega_0 r_0\,.
\end{equation}
Taylor expansion of the Lagrangian with respect to $x$ leads to
\begin{equation}
L \approx \frac{1}{2}\left[r_{\rm{0}} ^2\Omega_{\rm{0}}^{\rm{2}} - 
\Phi(r_{\rm{0}}) + \dot{x}^{\rm{2}} + \dot{y}^{\rm{2}} + 4\Omega_{\rm{0}} 
x\dot{y} + 4A\Omega_{\rm{0}} x^{\rm{2}} \right]\,,
\end{equation}
where $A$ denotes Oort's constant, $A = -\frac{1}{2}r_{\rm{0}} 
\left(\frac{d\Omega}{dr}\right)_{\rm{r_0}}$.
The equations of motion are then given by the standard epicyclic equations, 
\begin{equation}
\ddot{x} = 2\Omega_{\rm{0}} \dot{y} + 4A\Omega_{\rm{0}} x \quad {\rm and} \quad 
\ddot{y} = -2\Omega_{\rm{0}} \dot{x}\,,
\end{equation}
which can be solved straightforward in terms of linear and trigonometric
functions (cf. equations 10 and 11 below). The linear terms give the mean 
motion of the stars, whereas the trigonometric terms describe the epicyclic 
oscillations of the stars around the mean motion with a period of
$\kappa = \sqrt{-4\Omega_{\rm{0}} B}$,
where $B = A - \Omega_{\rm{0}}$ denotes Oort's second constant. The plane 
stellar orbits conserve two integrals of motion, one related to angular
momentum, 
\begin{equation}
J_{\rm{2}} = \frac{\partial L}{\partial \dot{y}} = \Omega_{\rm{0}} r_{\rm{0}} +
2 \Omega_{\rm{0}} x + \dot{y}\,,
\end{equation}
and the radial action integral,
\begin{equation}
J_{\rm{1}} = \frac{1}{2\kappa}\left(u^{\rm{2}} + \frac{\kappa^{\rm{2}}}
{4B^{\rm{2}}} v^{\rm 2} \right).
\end{equation}
The radial action integral is defined in terms of the radial velocity of a
star $u = \dot{x}$ and the circumferential velocity referred to the mean 
circular velocity of all stars at the position of the star, $ v = \dot{y} + 
2Ax = J_{\rm{2}} - \Omega_{\rm{0}} r_{\rm{0}} + 2Bx$. Next, the Hamiltonian of 
the stellar orbits is derived from the Lagrangian (4) and expressend in 
terms of the integrals of motion giving 
\begin{equation}
H = \kappa J_{\rm{1}} + \frac{A}{2B}\left(J_{\rm{2}} - \Omega_{\rm{0}}
r_{\rm{0}}\right)^{\rm{2}} - \frac{1}{2}\Omega_{\rm{0}}^{\rm{2}} 
r_{\rm{0}}^{\rm{2}}\,.
\end{equation}
Angle variables $w_{\rm{1}}$, $w_{\rm{2}} $ are introduced as canonical 
conjugate variables to the integrals of motion,
\begin{eqnarray}
\dot{w}_{\rm{1}} = \frac{\partial H}{\partial J_{\rm{1}}} = \kappa & , & 
\dot{w}_{\rm{2}} = \frac{\partial H}{\partial J_{\rm{2}}} = 
\frac{A}{B}\left(J_{\rm{2}} - \Omega_{\rm{0}} r_{\rm{0}} \right)\,.
\end{eqnarray}
Assembling the results from above, the orbit equations can be written as
\begin{eqnarray}
x & = & \frac{J_{\rm{2}} - \Omega_{\rm{0}} r_{\rm{0}}}{-2B} + 
\sqrt{\frac{2J_{\rm{1}}}{\kappa}}\sin{w_{\rm{1}}}\,, \nonumber\\  
y & = & w_{\rm{2}} - \frac{\sqrt{2\kappa
 J_{\rm{1}}}}{2B}\cos{w_{\rm{1}}}\,, \nonumber \\
u & = & \sqrt{2 \kappa J_{\rm{1}}} \cos{w_{\rm{1}}}\,,  \nonumber\\ 
v & = & \frac {2B}{\kappa}\sqrt{2\kappa J_{\rm{1}}}\sin{w_{\rm{1}}}\,.
\end{eqnarray}
These equations show that the mean motions of the stars are given by
\begin{eqnarray}
x & = &\frac{J_{\rm{2}} - \Omega_{\rm{0}} r_{\rm{0}}}{-2B} = const.\,,
\nonumber\\ y & = & \frac{A}{B}\left(J_{\rm{2}} - \Omega_{\rm{0}} r_{\rm{0}} 
\right)t = -2Axt\,,
\end{eqnarray}
describing the linear shear flow. The trigonometric terms describe the 
epicyclic oscillations around the mean motion, with their amplitude determined 
by the radial action integral $J_{\rm{1}}$.

\section{Boltzmann equation}

The ensemble of stars is described statistically by the distribution function
of stars in phase space $f$. Its time evolution is determined by the
collisionless Boltzmann equation
\begin{equation}
\frac{\partial f}{\partial t} + \left[f, H\right] = 0\,,
\end{equation}
where the Poisson bracket indicates the divergence of the Hamilton flow in
phase space. If integrals of motion are used the Poisson bracket has the 
form
\begin{equation}
\left[f, H\right] = \frac{\partial f}{\partial w_{\rm{1}}}\frac{\partial H}
{\partial J_{\rm{1}}} + \frac{\partial f}{\partial w_{\rm{2}}}
\frac{\partial H}{\partial J_{\rm{2}}}\,.
\end{equation}
Following Julian \& Toomre (1966), the Schwarzschild distribution,
\begin{equation}
f_0 = \frac{\kappa}{-2B}\frac{\Sigma_0}{2\pi\sigma_{\rm u}^2}
\exp{-\frac{\kappa J_1}{\sigma_{\rm u}^2}}
\end{equation}
is chosen as the distribution function of the homogenous background of the
sheet.
$\Sigma_{\rm{0}}$ denotes the surface density of the disk, which is kept 
constant over the sheet. $\sigma_{\rm{u}}$ is the radial velocity dispersion of
the stars. The circumferential velocity dispersion is given according to the
epicyclic ratio $\sigma_{\rm{v}} / \sigma_{\rm{u}} = 4 B^{\rm{2}}/
\kappa^{\rm{2}}$. Since the stationary Schwarzschild distribution depends only
on the radial action integral, it solves the Boltzmann equation by definition
(cf.~equation 13).

\section{Linear perturbation analysis}

Spiral arms are usually thought to be only minor perturbations of galactic 
disks. Thus I choose a perturbation `Ansatz' of the form
\begin{equation}
f = f_{\rm{0}} + \delta f \,, \; H = H_{\rm{0}} + \delta \Phi\,,
\end{equation}
where $H_{\rm{0}}$ is the Hamiltonian (8) of the unpertubed orbits. The
physical idea is to subject the shearing sheet to a small spiral--like potential
perturbation $\delta \Phi$ and to determine the disk response $\delta f$ by
solving the Boltzmann equation. The Boltzmann equation is linearized 
accordingly,
\begin{equation}
\frac{\partial \delta f}{\partial t} + \left[f_{\rm{0}} , \delta \Phi \right]
+ \left[ \delta f , H_{\rm{0}} \right] = 0 \,,
\end{equation}
where the quadratic Poisson bracket $[\delta f, \delta \Phi]$ has been
neglected. The role of the quadratic term will be investigated in a further
paper (Fuchs 2001, in preparation). A brief first report of that work was 
given in Fuchs (1991).
The potential perturbation is Fourier transformed with respect to the spatial 
coordinates and I consider a single Fourier component
\begin{equation}
\delta \Phi_{\rm{\bf{k}}} = \Phi_{\rm{\bf{k}}} \exp{i\left(k_{\rm{x}} x +
 k_{\rm{y}} y\right)}
\end{equation}
in equation (16). The Poisson brackets have then the explicit form
\begin{equation}
\left[f_{\rm{0}}, \delta\Phi_{\rm{\bf{k}}} \right] = \frac{\sqrt{2\kappa 
J_{\rm{1}}}}{\sigma_{\rm{u}}^{\rm{2}}} \left[i k_{\rm{x}} \cos{w_{\rm{1}}}
+ \frac{\kappa}{2B}i k_{\rm{y}} \sin{w_{\rm{1}}} \right] f_{\rm{0}} \delta 
\Phi_{\rm{\bf{k}}}
\end{equation}
and
\begin{equation}
\left[ \delta f, H_{\rm{0}} \right] = \kappa \frac{\partial \delta f}{\partial 
w_{\rm{1}}} + \frac{A}{B}\left(J_{\rm{2}} - \Omega_{\rm{0}} r_{\rm{0}} \right)
\frac{\partial \delta f}{\partial w_{\rm{2}}},
\end{equation}
which leads to the linearized Boltzmann equation
\begin{eqnarray}
&&
\frac{\partial \delta f}{\partial t} + \kappa \frac{\partial \delta f}{\partial
w_{\rm{1}}} + \frac{A}{B} \left(J_{\rm{2}} - \Omega_{\rm{0}} r_{\rm{0}} \right) 
\frac{\partial \delta f}{\partial w_{\rm{2}}}  \nonumber\\
&&+ \frac{\sqrt{2 \kappa J_{\rm{1}}}}{\sigma_{\rm{u}}^{\rm{2}}} \Big[i 
k_{\rm{x}} \cos{w_{\rm{1}}} + \frac{\kappa}{2B} i k_{\rm{y}} \sin{w_{\rm{1}}}
\Big] \frac{\kappa}{-2B} \nonumber \\ && \cdot \frac{\Sigma_{\rm{0}}}{2 \pi 
\sigma_{\rm{u}}^{\rm{2}}} \exp{\left(-\frac{\kappa J_{\rm{1}}}
{\sigma_{\rm{u}}^{\rm{2}}}\right)} \Phi_{\rm{\bf{k}}} \exp{\left( i k_{\rm{x}} x
 + i k_{\rm{y}} y \right)} = 0\,. 
\end{eqnarray}
When $x$ and $y$ are substituted from equations (10), equation (20) is a 
linear inhomogenous partial differential equation with $t$, $w_{\rm{1}}$,
and $w_{\rm{2}}$ as independent variables. It becomes immediately clear 
that the coefficients of the equation do not depend on the time coordinate $t$
and the angle variable $w_{\rm{2}}$. This allows a Fourier transformation of 
equation (20) with respect to $t$ and $w_{\rm{2}}$, which may be written in 
abbreviated form as
\begin{eqnarray}
&& i\omega\delta f_{\rm{\omega ,l}} + \kappa \frac{d \delta
 f_{\rm{\omega ,l}}}{d w_{\rm{1}}} + \frac{A}{B}\left(J_{\rm{2}}
 - \Omega_{\rm{0}} r_{\rm{0}} \right) i l \delta f_{\rm{\omega ,l}} 
\nonumber\\
&& = e^{i l w_{\rm 2}} \int^{\infty}_{-\infty} dw'_{\rm{2}}
e^{-i l w'_{\rm 2}} {Inhom}_{\rm{\omega}}\left(J_{\rm{1}}, J_{\rm{2}},
w_{\rm{1}} \right) e^{i k_y w'_{\rm 2}} \nonumber \\ 
&& =  2\pi e^{i l w_{\rm 2}} {Inhom}_\omega \left(J_{\rm{1}},
J_{\rm{2}},w_{\rm{1}} \right) \delta \left(k_{\rm{y}} - l\right)\,,
\end{eqnarray}
where $\delta (k_{\rm y} - l)$ denotes a delta function.
Equation (20) will be solved explicitely below. However, it becomes clear 
from equation (21) that
the solution $f_{\rm{\omega ,l}}$ must be proportional to $ \Phi_{\rm{\bf{k}}}
e^{i l w_{\rm 2}}\delta\left(k_{\rm{y}} - l\right)$, because of the 
linearity of the equation and the consideration that only a non--zero
perturbation leads to a disk response. Transforming back to $w_{\rm{2}}$ space, 
this implies a functional dependence of $\delta f$ on $w_{\rm 2}$ as
\begin{equation}
\delta f \propto e^{i k_y w_{\rm 2}}.
\end{equation}
Furthermore it is advantageous to split off the background distribution 
function from the disk response, $\delta f_{\rm{\omega}} = f_{\rm{1, \omega}}
f_{\rm{0}}$, and I obtain the ordinary differential equation
\begin{eqnarray}
&& i \omega f_{\rm{1,\omega}} + \kappa \frac{d f_{\rm{1,\omega}}}
{d w_{\rm{1}}} + ik_{\rm{y}} \frac{A}{B}\left( J_{\rm{2}} - 
\Omega_{\rm{0}} r_{\rm{0}} \right) f_{\rm{1,\omega}} \nonumber\\ 
&& + \frac{\sqrt{2\kappa J_{\rm{1}}}}{\sigma_{\rm{u}}^{\rm{2}}}
\left[ik_{\rm{x}} \cos{w_{\rm{1}}} + \frac{\kappa}{2B}ik_{\rm{y}} 
\sin{w_{\rm{1}}} \right] \Phi_{\rm{\bf{k},\omega}} \nonumber\\
&& \cdot  {\rm exp} \{ ik_{\rm{x}} \left[ \frac{J_{\rm{2}} - \Omega_{\rm{0}} r_{
\rm{0}}}{-2B} + \sqrt{\frac{2J_{\rm{1}}}{\kappa}}\sin{w_{\rm{1}}} \right]       
\nonumber \\
&& - ik_{\rm{y}}\frac{\sqrt{2\kappa J_{\rm{1}}}}{2B}\cos{w_{\rm{1}}} \} =
 0\,. 
\end{eqnarray}
Equation (23) can be solved by `variation of the constant'. Very similar
equations have been treated with the same method by Lin, Yuan, \& Shu (1969) 
and Julian \& Toomre (1966) and I refer in particular to the lecture notes by
Lin (1970) for a full explanation of the technique. The homogenous part of 
equation (23) has solutions
\begin{equation}
f_{\rm{1,\omega}} \propto \exp{-\frac{i}{\kappa}\left(\omega + \frac{A}{B}
k_{\rm{y}} \left( J_{\rm{2}} - \Omega_{\rm{0}} r_{\rm{0}} \right) \right) 
w_{\rm{1}}}\,,
\end{equation}
and `variation of the constant' means to look for a particular solution of
equation (23) in the form
\begin{equation}
f_{\rm{1,\omega}} =  C(w_{\rm{1}}) \exp{-\frac{i}{\kappa}\left(
\omega + \frac{A}{B}k_{\rm{y}}\left(J_{\rm{2}} - \Omega_{\rm{0}} r_{\rm{0}}
\right)\right)w_{\rm{1}}}\,.
\end{equation}
Inserting this into equation (23) and solving for $C(w_{\rm{1}})$ gives
\begin{eqnarray}
C(w_{\rm{1}}) & = & - \frac{i\xi}{\sigma^{\rm{2}}_{\rm{u}}}\exp{\left(ik_{\rm{x}
} \frac{J_{\rm{2}} - \Omega_{\rm{0}} r_{\rm{0}}}{-2B}\right)} \Phi_{\rm{\bf{k},
\omega}} \exp{i\eta\overline{w}}\nonumber\\
 &\cdot &\int^{w_{\rm 1} - \overline{w}}_0 dw'_{\rm{1}} \cos{w'_{\rm{1}}}
 \exp{i \left[\xi \sin{w_{\rm{1}}} + \eta w'_{\rm{1}}\right]}\,,
\end{eqnarray}
where the abbreviations $\xi = \sqrt{\frac{2J_{\rm{1}}}{\kappa}}
\sqrt{k^{\rm{2}}_{\rm{x}} + \frac{\kappa^{\rm{2}}}{4B^2}k^{\rm{2}}_{\rm{y}}}$ 
and $\eta = \frac{1}{\kappa}\left(\omega + k_{\rm{y}} \frac{A}{B}\left
(J_{\rm{2}} - \Omega_{\rm{0}} r_{\rm{0}} \right)\right)$
have been introduced to keep the formulae compact. $\overline{w}$ is defined as
$\overline{w} = \arctan{\left(\frac{\kappa}{2B}\frac{k_{\rm{y}}}{k_{\rm{x}}}
\right)}$. The general solution of equation (23) is then given by the
particular integral (25) plus the solution of the homogenous equation
multiplied by an arbitrary integration constant,
\begin{eqnarray}
&&f_{\rm{1,\omega}} = \frac{\Phi_{{\rm{\bf k}},\omega}}
 {\sigma^{\rm{2}}_{\rm{u}}} \exp{\left(ik_{\rm{x}}\frac{J_{\rm{2}} - 
 \Omega_{\rm{0}} r_{\rm{0}}}{-2B}\right)} \{D - i \xi \exp{ i \eta 
 \overline{w}} \\
&& \cdot \int^{w_1 - \overline{w}}_0 dw'_{\rm{1}} \cos{w'_{\rm{1}}} 
 \exp{i \left[\xi \sin{w'_{\rm{1}}} + \eta w'_{\rm{1}} \right]} \}
 \exp{-i\eta w_{\rm{1}}}\,. \nonumber
\end{eqnarray}
The integration constant $D$ is determined by the requirement that the 
distribution function must be defined in a unique way in velocity space, 
i.e.~must be periodic with respect to the angle variable $w_{\rm{1}}$ with a
period of $2\pi$,
\begin{equation}
f_{\rm{1,\omega}}\left(w_{\rm{1}}\right) = f_{\rm{1,\omega}}\left(w_{\rm{1}} + 
2\pi\right).
\end{equation}
Some algebra as indicated in Lin (1970) leads to the final result
\begin{eqnarray}
&& f_{\rm{1,\omega}}  = \nonumber \\ && 
-\frac{1}{\sigma^{\rm{2}}_{\rm{u}}}
\exp{\left(i k_{\rm{x}} \frac{J_{\rm{2}} - \Omega_{\rm{0}} r_{\rm{0}}}{-2B}
\right)} \Phi_{\rm{\bf{k},\omega}} \exp{(i \xi \sin{\left(w_{\rm{1}} -      
\overline{w}\right)})} \nonumber\\ &&                                             
 \cdot \{ 1 - \frac{\eta}{2 \sin{(\pi\eta)}} \int^{+\pi}_{-\pi} dw'_{\rm{1}} 
{\rm exp } i \big[ \eta w'_{\rm{1}} - \xi \sin{w'_{\rm{1}}} \nonumber \\ &&
  \cdot  \cos{\left(w_{\rm{1}} - \overline{w}\right)}- \xi \left( 1 + 
 \cos{w'_{\rm{1}}} \right) \sin{\left( w_{\rm{1}} - \overline{w}\right)}
 \big] \}\,.
\end{eqnarray}
The disk response to the single Fourier component of the potential perturbation
is thus given by
\begin{equation}
\delta f = f_{\rm{0}} f_{\rm{1,\omega}} \exp{\left(i \omega t + i k_{\rm{y}}
w_{\rm{2}}\right)}\,.
\end{equation}
Note that according to equations (10) and the definitions of $\xi$ and
$\overline{w}$
\begin{equation}
k_{\rm{x}} \frac{J_{\rm{2}} - \Omega_{\rm{0}} r_{\rm{0}}}{-2B} + \xi
\sin{\left(w_{\rm{1}} - \overline{w}\right)} + k_{\rm{y}}w_2  = k_{\rm{x}}
x + k_{\rm{y}} y\,.
\end{equation}
This shows that the disk response follows to a first approximation the spatial
pattern of the potential perturbation. However, the second term in the curly
brackets in equation (29) depends through $\eta$ on
$(J_{\rm{2}} - \Omega_{\rm{0}} r_{\rm{0}} )$ and thus on the spatial coordinate
$x$, so that contrary to $\Phi_{\rm{k}, \omega}$, $f_{\rm{1},\omega}$ is 
{\em  not } a Fourier component of a Fourier decomposition of the disk response.

\section{Self consistent perturbations}                                         

The shearing sheet is assumed to be self-gravitating. Thus the potential
perturbation and the corresponding disk response have to satisfy the Poisson
equation,
\begin{equation}
\{ \frac{\partial}{\partial x^{\rm{2}}} + \frac{\partial}{\partial
 y^{\rm{2}}} + \frac{\partial}{\partial z^{\rm{2}}} \} \delta \Phi 
 = 4 \pi G \delta \Sigma \delta\left(z\right)\,,
\end{equation}
where $\delta \Sigma$ denotes the surface density of the disk response and
$\delta (z)$ is a delta function with respect to the vertical spatial coordinate
$z$. $G$ is the constant of gravity. Background potentials do not 
enter into equation (32) because of the linearity of the Poisson equation. Off 
the midplane the right hand side of equation (32) is equal to zero. This leads
immediately to a three--dimensional extension of a potential perturbation (17)
in the form 
\begin{equation}
\delta\Phi_{\rm \bf{k}} = \Phi_{\rm{\bf{k}}} \exp \left[ i\left(k_{\rm{x}} x +
k_{\rm{y}} y\right) - \sqrt{k_{\rm{x}}^{\rm{2}} + k_{\rm{y}}^{\rm{2}}}\,|z| 
\right]\,.
\end{equation}
Integrating both sides of the Poisson equation with respect to $z$ gives the 
surface density required to sustain the potential (33),
\begin{eqnarray}
&&\int^{+\infty}_{-\infty} dz \{ {\frac{\partial ^{\rm{2}}}{\partial
x^{\rm{2}}} + \frac{\partial ^{\rm{2}}}{\partial y^{\rm{2}}} + 
\frac{\partial ^{\rm{2}}}{\partial z^{\rm{2}}}} \} \Phi_{\rm{\bf{k}}}
\nonumber\\
&&\cdot \exp{\left[i\left(k_{\rm{x}} x + k_{\rm{y}} y \right) - 
\sqrt{k^{\rm{2}}_{\rm{x}} + k^{\rm{2}}_{\rm{y}}}\,|z|\right]}\nonumber\\
&&= -2\sqrt{k^{\rm{2}}_{\rm{x}} + k^{\rm{2}}_{\rm{y}}}\, \Phi_{\rm{\bf{k}}}
\exp i \left(k_{\rm{x}} x + k_{\rm{y}} y \right)\nonumber\\
&&= 4\pi G \Sigma_{\rm{\bf{k}}} \exp{i \left(k_{\rm{x}} x + k_{\rm{y}} y
\right)}\,,
\end{eqnarray}
where $\Sigma_{\rm{\bf{k}}}$ denotes the Fourier coefficients of the Fourier
transform of $\delta \Sigma$.

It was pointed out in the previous section that the disk response to a single 
Fourier component of a general potential perturbation as considered in 
equations (17) and (33) is not a Fourier component of the disk response to the
general potential perturbation. Kalnajs (1971) has shown a way to overcome this
difficulty by taking Fourier transforms of both sides of the Poisson equation 
in the following way
\begin{eqnarray}
&&\frac{1}{\left(2\pi\right)^2}\int^{+\infty}_{-\infty} dz 
\int^{+\infty}_{-\infty} dx \int^{+\infty}_{-\infty} dy
e^{-i\left(k'_x x + k'_y y \right)}\nonumber\\
&&\{\frac{\partial^{\rm{2}}}{\partial x^{\rm{2}}} +\frac{\partial^{\rm{2}}}
{\partial y^{\rm{2}}} +\frac{\partial^{\rm{2}}}{\partial z^{\rm{2}}} \}
\int^{+\infty}_{-\infty}dk_{\rm{x}}\int^{+\infty}_{-\infty}dk_{\rm{y}} 
\Phi_{\rm{\bf{k}}}\nonumber \\
&&\cdot \exp\left[i\left(k_{\rm{x}} x + k_{\rm{y}} y\right) - 
\sqrt{k^{\rm{2}}_{\rm{x}} + k^{\rm{2}}_{\rm{y}}}\,|z|\right] \nonumber\\
&& = \frac{4\pi G}{\left(2\pi\right)^{\rm{2}}} \int^{+\infty}_{-\infty}dx
\int^{+\infty}_{-\infty}dy e^{-i\left(k'_x x + k'_y y \right)} \nonumber \\
&&\int^{+\infty}_{-\infty}dk_{\rm{x}} \int^{+\infty}_{-\infty}dk_{\rm{y}} 
\int^{+\infty}_{-\infty}du \int^{+\infty}_{-\infty}dv f_{\rm{0}} 
f_{\rm{1,\omega}} e^{ik_y w_2}\,.
\end{eqnarray}
The $k_{\rm{x}} , k_{\rm{y}}$ integrations indicate that all Fourier 
coefficients have to be considered together, while the integrations of the 
distribution function of the disk response over the $u$ und $v$ velocity 
components on the right hand side of euation (35) give the surface density of
the disk response. The left hand side of equation (35) can be evaluated by 
carrying out first the quadrature with respect to $z$, giving
\begin{eqnarray}
-\frac{2}{\left(2\pi\right)^{\rm{2}}} \int^{+\infty}_{-\infty}dx 
\int^{+\infty}_{-\infty}dy \int^{+\infty}_{-\infty}dk_{\rm{x}} 
\int^{+\infty}_{-\infty}dk_{\rm{y}} \nonumber\\ \cdot
\sqrt{k_{\rm{x}}^{\rm{2}} + k_{\rm{y}}^{\rm{2}}}\,\Phi_{\rm{\bf{k}}}
\exp{i \left[\left(k_{\rm{x}} -k'_{\rm{x}}\right) x + 
\left(k_{\rm{y}} - k'_{\rm{y}} \right) y \right]} \nonumber\\
= -2 \sqrt{{k'_{\rm{x}}}^{\rm{2}} + {k'_{\rm{y}}}^{\rm{2}}}\,\Phi_{\rm{\bf{k'}}}
\,.
\end{eqnarray}
Equation (35) represents in effect an integral equation which has to be solved 
to obtain self consistent pairs of potential and density perturbations of the 
sheet.

\section{Derivation of the Volterra integral equation}

In order to derive the central integral equation of this study the right hand 
side of equation (35) has to be evaluated explicitly. The multiple integrals 
imply an integration over the entire phase space and it is convenient to change
the integrations from spatial $x$, $y$ and $u$, $v$ velocity coordinates to 
$J_1$, $J_2$ and $w_1$, $w_2$ variables. Using equations (10) it can be shown 
that the Jacobian of the transformation is equal to one,
\begin{equation}
\left|\frac{\partial \left(x,y,u,v\right)}{\partial \left(J_{\rm{1}} ,J_{\rm{2}}
,w_{\rm{1}} ,w_{\rm{2}} \right)}\right| = 1\,.
\end{equation}
Inserting $f_{\rm{0}}$, $f_{\rm{1,\omega}}$, $x$, and $y$ into the right hand
side of equation (35) leads to
\begin{eqnarray}
&&-\frac{4\pi G}{\left(2\pi\right)^{\rm{2}}} \int^{2\pi}_{0}dw_{\rm{1}}
\int^{+\infty}_{-\infty} dw_{\rm{2}}\int^{\infty}_0 d J_{\rm{1}}
\int^{+\infty}_{-\infty} dJ_{\rm{2}} \frac{\kappa}{-2B} \nonumber \\
&& \cdot \frac{\Sigma_{\rm{0}}}{2\pi\sigma_{\rm{u}}^{\rm{2}}} 
\exp{-\left(\frac{\kappa}{\sigma_{\rm{u}}^{\rm{2}}}J_{\rm{1}}\right)}
\int^{+\infty}_{-\infty} dk_{\rm{x}}\int^{+\infty}_{-\infty} dk_{\rm{y}}
\frac{\Phi_{\rm{\bf{k}}}}{\sigma_{\rm{u}}^{\rm{2}}} \nonumber \\ && \cdot
\exp{ i \left[ k_{\rm{x}} \frac{J_{\rm{2}} - \Omega_{\rm{0}} r_{\rm{0}}}{-2B} + 
k_{\rm{y}}w_{\rm{2}} + \xi\sin{\left(w_{\rm{1}} - \overline{w}\right)}\right]}
\nonumber \\ && \cdot
\{ 1-\frac{\eta}{2\sin{(\pi\eta})}\int^{+\infty}_{-\infty}
dw'_{\rm{1}} {\rm exp} i [ \eta w'_{\rm{1}} - \xi\sin{w'_{\rm{1}}} \nonumber \\
&& \cdot \cos{\left(w_{\rm{1}} - 
\overline{w}\right)}-\xi\left(1 + \cos{w'_{\rm{1}}}\right)\sin{\left(w_{\rm{1}}
- \overline{w}\right)}]\} \nonumber \\
&&\cdot {\rm exp} -i \Big[ k'_{\rm{x}} \frac{J_{\rm{2}} - \Omega_{\rm{0}} 
r_{\rm{0}}}{-2B} + k'_{\rm{x}} \sqrt{\frac{2J_{\rm{1}}}{\kappa}}{\rm sin} 
w_{\rm{1}} \nonumber \\
&& + k'_{\rm{y}} w_{\rm{2}} - k'_{\rm{y}} \frac{\sqrt{2\kappa J_{\rm{1}}}}{2B}
\cos{w_{\rm{1}}}\Big]\,.
\end{eqnarray}
All of the quadratures but one are, even if tedious, straightforward. The
integration over the angle variable $w_{\rm{2}}$ is particularly simple,
\begin{equation}
\frac{1}{2\pi}\int^{+\infty}_{-\infty}dw_{\rm{2}} e^{i\left(k_{\rm y} - 
k'_{\rm y} \right)w_2} = \delta\left(k_{\rm{y}} - k'_{\rm{y}}\right)\,,
\end{equation}
and the integration over $k_{\rm{y}}$ contracts the integrand of (38) to 
$k_{\rm{y}} = k'_{\rm{y}}$. This shows that the integral equation (35) is 
separating in the circumferential wave numbers $k_{\rm{y}}$.

The integrand of equation (38) is split according to the two terms in the 
curly bracket into two parts. The $w_{\rm{1}}$ integration of the first
part of the integrand leads to a quadrature
\begin{eqnarray}
\int^{2\pi}_{0}d w_{\rm{1}} \exp i\left[\xi\sin\left(w_{\rm{1}} - 
\overline{w}\right) - \xi'\sin\left(w_{\rm{1}} -
 \overline{w}'\right)\right]\nonumber\\
=\int^{2\pi}_{0}d w_{\rm{1}} \exp i [\left(\xi^{\rm{2}} -
 2\xi\xi' \cos\left(\overline{w} - \overline{w}' \right)+ 
 {\xi'}^{\rm{2}}\right)^{\frac{1}{2}} \nonumber \\  \cdot
\sin\left(w_{\rm{1}} - \overline{\overline{w}}\right)]\,,
\end{eqnarray}
where $\xi'$ and $\overline{w}'$ are defined analogous to $\xi$ 
and $\overline{w}$ for primed wave numbers. $\overline{\overline{w}}$ 
is given by \\ $\overline{\overline{w}} = 
{\rm arctan}\left(\left(\xi\sin\overline{w} - \xi'\sin\overline{w}'\right)
/\left(\xi\cos\overline{w} - \xi'\cos\overline{w}'\right)\right)$.

A Fourier expansion of the integrand of (40) with respect to $w_1$ 
(Abramowitz \& Stegun 1968, section 9.1 ) gives immediately the result of the 
quadrature
\begin{equation}
2\pi J_{\rm{0}}\left(\left(\xi^{\rm{2}} - 2\xi\xi'
\cos\left(\overline{w} - \overline{w}'\right) + 
\xi'^{\rm{2}}\right)^{\rm{\frac{1}{2}}}\right)\,,
\end{equation}
where $J_{\rm{0}}$ denotes the Bessel function of the first kind. The second 
part of the integrand of (38) is slightly more complicated,
\begin{eqnarray}
&&\int^{2\pi}_{0}d w_{\rm{1}} \exp i \left[-\xi\sin\left(w_{\rm{1}}'
+ w_{\rm{1}} - \overline{w}\right) - \xi' 
\sin\left(w_{\rm{1}} - \overline{w}'\right)\right]
\nonumber\\ &&
=2\pi J_{\rm{0}} \left(\left(\xi^{\rm{2}} + 2\xi\xi'
\cos\left(w'_{\rm{1}} - \overline{w} + \overline{w}'\right) + 
\xi'^{\rm{2}} \right)^{\frac{1}{2}}\right)\,.
\end{eqnarray}
I consider next the integration over the action variable $J_{\rm{1}}$ which 
leads for the first part of the integrand of (38) to a quadrature
\begin{eqnarray}
\int^{\infty}_{0}dJ_{\rm{1}} \exp -\left(\frac{\kappa}
{\sigma_{\rm{u}}^{\rm{2}}}J_{\rm{1}}\right)2\pi J_{\rm{0}} 
\Big(\Big[\frac{2}{\kappa}\left(k_{\rm{x}}^{\rm{2}} + \frac{\kappa^2}
{4B^2}k_{\rm{y}}^{\rm{2}}\right) \nonumber\\
+ \frac{2}{\kappa}\left(k_{\rm{x}}'^{\rm{2}} + \frac{\kappa^2}{4B^2}
k_{\rm{y}}'^{\rm{2}}\right)-\frac{4}{\kappa}\sqrt{k_{\rm{x}}^{\rm{2}} + 
\frac{\kappa^{\rm{2}}}{4B^{\rm{2}}}k_{\rm{y}}^{\rm{2}}} \nonumber\\
\cdot \sqrt{k_{\rm{x}}'^{\rm{2}} + \frac{\kappa^{\rm{2}}}{4B^{\rm{2}}}
k_{\rm{y}}'^{\rm{2}}}\cos\left(\overline{w} - \overline{w}'\right) 
\Big] ^{\rm{\frac{1}{2}}}\sqrt{J_{\rm{1}}}\Big)\,,
\end{eqnarray}
if $\xi$ and $\xi'$ are inserted into the argument of the Bessel function.
According to formula (6.631) of Gradshteyn \& Ryzhik (1965) this can be 
evaluated as
\begin{eqnarray}
&&\frac{2\pi\sigma_{\rm{u}}^{\rm{2}}}{\kappa}
{\rm exp} -\frac{\sigma_{\rm{u}}^{\rm{2}}}{\kappa^{\rm{2}}}
\Big[\frac{1}{2}\left(k_{\rm{x}}^{\rm{2}} + \frac{\kappa^{\rm{2}}}
{4B^{\rm{2}}}k_{\rm{y}}^{\rm{2}}\right)
+\frac{1}{2}\left({k'}_{\rm{x}}^{\rm{2}} + \frac{\kappa^{\rm{2}}}{4B^{\rm{2}}}
{k'}_{\rm{y}}^{\rm{2}}\right)\nonumber\\ &&
-\sqrt{k_{\rm{x}}^{\rm{2}} + \frac{\kappa^{\rm{2}}}{4B^{\rm{2}}}
k_{\rm{y}}^{\rm{2}}}\,\sqrt{{k'}_{\rm{x}}^{\rm{2}} + \frac{\kappa^{\rm{2}}}
{4B^{\rm{2}}}{k'}_{\rm{y}}^{\rm{2}}}\,\cos\left(\overline{w} - 
\overline{w}'\right)\Big]\,,
\end{eqnarray}
and similarly for the second part of the integrand of (38)
\begin{eqnarray}
&&\frac{2\pi\sigma_{\rm{u}}^{\rm{2}}}{\kappa}
{\rm exp} -\frac{\sigma_{\rm{u}}^{\rm{2}}}{\kappa^{\rm{2}}}
\Big[\frac{1}{2}\left(k_{\rm{x}}^{\rm{2}} + \frac{\kappa^{\rm{2}}}
{4B^{\rm{2}}}k_{\rm{y}}^{\rm{2}}\right)
+\frac{1}{2}\left({k'}_{\rm{x}}^{\rm{2}} + \frac{\kappa^{\rm{2}}}{4B^{\rm{2}}}
{k'}_{\rm{y}}^{\rm{2}}\right) \nonumber\\ &&
-\sqrt{k_{\rm{x}}^{\rm{2}} + \frac{\kappa^{\rm{2}}}{4B^{\rm{2}}}
k_{\rm{y}}^{\rm{2}}}\,\sqrt{{k'}_{\rm{x}}^{\rm{2}} + \frac{\kappa^{\rm{2}}}
{4B^{\rm{2}}}{k'}_{\rm{y}}^{\rm{2}}} \nonumber \\ && \cdot
\cos\left( w'_{\rm 1} -\overline{w} + \overline{w}'\right)\Big]\,.
\end{eqnarray}

In the second part of the integrand of (38) a quadrature over the auxiliary 
angle variable $w_{\rm{1}}'$ has to be carried out,
\begin{equation}
\int^{+\pi}_{-\pi}dw'_{\rm{1}} \exp{i \left[\eta w'_1-\Lambda
\cos\left(w'_1 - \overline{w} + \overline{w}'\right)\right]}\,,
\end{equation}
where I have introduced $\Lambda$ as an abbreviation for
$\Lambda = \frac{\sigma_u^2}{\kappa^2}\sqrt{k_{\rm x}^2 + \frac{\kappa^2}{4B^2}
k_{\rm y}^2}\sqrt{{k'}_{\rm x}^2 + \frac{\kappa^2}{4B^2}{k'}_{\rm y}^2}$. A
Fourier transform with respect to $w'_1$ gives (Abramowitz \& Stegun, 1968,
Section 9.6)
\begin{eqnarray}
&&\int^{+\pi}_{-\pi}dw'_1\left(\cos(\eta w'_1) +i\sin(\eta w'_1)
\right)\nonumber\\
&& \cdot\left[I_0 \left(-\Lambda\right)+2\sum_{n=1}^{\infty}I_{\rm n} 
\left(-\Lambda\right)
\cos (n\left(w'_1 - \overline{w}+\overline{w}'\right))\right]
\nonumber\\ &&
=I_0 \left(\Lambda\right)\frac{2\sin (\pi \eta)}{\eta}\nonumber\\ &&
+2\sum_{n=1}^{\infty}I_{\rm n} \left(-\Lambda\right)\left(-1\right)^{\rm n} 
\sin\left(\pi\eta\right)\cos\left(n\left(\overline{w}-\overline{w}'\right)
\right)\nonumber \\ && \cdot
\left[\frac{1}{\eta-n}+\frac{1}{\eta+n}\right]\nonumber\\ &&
+2 i \sum_{n=1}^{\infty}I_{\rm n} \left(-\Lambda\right)\left(-1\right)^{\rm n} 
\sin\left(\pi\eta\right)\sin\left(n\left(\overline{w}-\overline{w}'\right)
\right) \nonumber \\ && \cdot
\left[\frac{1}{\eta-n}-\frac{1}{\eta+n}\right]\,,
\end{eqnarray}
where the $I_{\rm n}$ denote modified Bessel functions of the first kind.

This leaves finally the integration over the variable $J_2$. The first
part of the integrand of (38) is evaluated easily,
\begin{eqnarray}
&&\int^{+\infty}_{-\infty}dJ_2 \exp \left[i\left(k_{\rm x} - k'{\rm _x}
\right)\frac{J_2 - \Omega_0 r_0}{-2B}\right]\nonumber\\ &&
 = 2\pi\left(-2B\right)\exp \left[i\left(k_{\rm x} - k'{\rm _x}
 \right)\frac{\Omega_0 r_0}{2B}\right]\delta\left(k_{\rm x} - k'_{\rm x}
 \right)\,.
\end{eqnarray}
The integration over $k_{\rm x}$ contracts the first part of the integrand of 
(38) to $k_{\rm x} = k'_{\rm x}$.

However the second part of the integrand needs more care. The first term of 
the series (47) leads to
\begin{eqnarray}
&&\int^{+\infty}_{-\infty}dJ_2 \exp\left[i\left(k_{\rm x} - k'{\rm _x}
\right)\frac{J_2 - \Omega_0 r_0}{-2B}\right]\left(-1\right)I_0 \left(\Lambda
\right)  \\ &&
=-2\pi \left(-2B\right)\exp\left[i\left(k_{\rm x} - k'_{\rm x}
\right)\frac{\Omega_0 r_0}{2B}\right]I_0\left(\Lambda\right)\delta 
\left(k_{\rm x} - k'_{\rm x}\right)\,,\nonumber
\end{eqnarray}
and the integration over $k_{\rm x}$ contracts this part of the second half of
the integrand again to $k_{\rm x} = k'_{\rm x}$. The next terms of the series
(47) lead to integrals of the form
\begin{eqnarray}
&&\int^{+\infty}_{-\infty} dJ_2 \exp\left[i\left(k_{\rm x} - k'_{\rm x}\right)
\frac{J_2 - \Omega_0 r_0}{-2B}\right] \nonumber \\ && \cdot
\left(-\eta\right)\left[\frac{1}
{\eta - n}\pm\frac{1}{\eta + n}\right]\,.
\end{eqnarray}
The auxiliary variable $\eta$ is substituted for the variable $J_2$ 
and we have $d\eta = \frac{A}{B}\frac{k'_{\rm y}}{\kappa}dJ_2$. Since the Oort 
constant $B$ is negative, I consider first the case of negative wave numbers 
$k'_{\rm y}$. The integral (50) is of the type
\begin{equation}
\lim_{\lambda\to 0}\int^{+\infty + \frac{i\lambda}{\kappa}}_{-\infty + 
\frac{i\lambda}{\kappa}}d\eta e^{i\Delta\eta}\frac{\eta}{\eta\pm n}\,,
\end{equation}
where $\Delta$ denotes $\Delta = \kappa\left(k'_{\rm x} - 
k_{\rm x}\right)/(2Ak'_{\rm y})$. According to its definition the variable
$\eta$ is related to the frequency $\omega$. I have assumed in the Fourier 
transforms throughout the previous sections a real frequency $\omega$. However,
according to Landau's rule this must be understood as a limiting value of
frequencies with negative imaginary parts $\lambda$. The reasoning is that
only perturbations, which {\em grow} in time from minus infinity are physically
relevant, and, if perturbations with other imaginary parts are considered,
these must be analytical continuations of the former (cf.~Binney \& Tremaine 
1987). According to the definition of the Fourier transform used here the
functional dependence of the perturbations is $e^{i\omega t} \propto 
e^{-\lambda t}$, so that negative imaginary parts $\lambda$ indicate
growing perturbations. This implies that in the integral (51) the integration 
contour must be chosen in that way that it encompasses the poles of the 
integrand in the complex domain of $\eta$ below the real axis. With these
caveats in mind the integral (51) can be evaluated as
\begin{eqnarray}
&&\lim_{\lambda\to 0}\int^{+\infty + \frac{i\lambda}{\kappa}}_{-\infty +
\frac{i\lambda}{\kappa}}d\eta' e^{i\Delta\left(\eta'\mp n \right)}
\frac{\eta' \mp n}{\eta'} \nonumber \\
&&= 2\pi e^{\mp i\Delta n}\delta\left(\Delta\right) \mp ne^{\mp i\Delta n}
\lim_{\lambda\to 0}\int^{+\infty + \frac{i\lambda}{\kappa}}_{-\infty + 
\frac{i\lambda}{\kappa}}d\eta' \frac{e^{i\Delta\eta'}}{\eta'} \nonumber \\ 
&&= 2\pi e^{\mp i\Delta n}\delta\left(\Delta\right) \mp n e^{\mp i\Delta n}
\nonumber \\ && \cdot
\left(i \pi + i{\cal P} \int^{+\infty}_{-\infty}d \eta'  \frac{\sin\left(
\Delta\eta'\right)}{\eta'}\right)\,,
\end{eqnarray}
where $\eta'$ has been substituted for $\eta' = \eta \pm n$.
$\cal P $ indicates the Cauchy principal value of the integral, and if 
the latter is evaluated as $\pi i\,{\rm sign}\left(\Delta\right)$ (Gradshteyn \&
Ryzhik 1965, equation 3.722), the integral (51) is equal to
\begin{equation}
2\pi e^{\mp i\Delta n}\left(\delta\left(\Delta\right) \mp                       
 i\,n\,u\left(\Delta\right)\right)\,,
\end{equation}
where $u\left(\Delta\right)$  denotes the unit step function,
\begin{equation}
u\left(\Delta\right) = \left \{ \begin{array}{cc}1, & \Delta >0 
\\ \frac{1}{2}, & \Delta=0 \\ 0, & \Delta<0 \end{array} \,. \right.
\end{equation}
Inserting this back into equation (50) gives
\begin{eqnarray}
&&- 2 \pi \frac{B}{A} \frac{\kappa}{k'_{\rm y}} \exp{-\left(
\frac{i\omega}{\kappa}\Delta\right)}\nonumber \\ &&\cdot
 \left[ \begin{array}{cc} \delta(\Delta) \cos(n\Delta)& - n \sin(n\Delta)
 u(\Delta) \\  0& +i\,n \cos(n\Delta) u(\Delta)
 \end{array} \right ]\,, 
\end{eqnarray}
where the upper row refers to the plus sign and the lower to the minus sign in
equation (50), respectively. The case of positive wave numbers $k'_{\rm y}$
can be treated in an analogous way, leading to a sign reversal,
\begin{eqnarray}
&&+ 2 \pi \frac{B}{A} \frac{\kappa}{k'_{\rm y}} \exp{-\left(
\frac{i\omega}{\kappa}\Delta\right)}\nonumber \\ &&\cdot
 \left[ \begin{array}{cc} \delta(\Delta) \cos(n\Delta)& - n \sin(n\Delta)
 u(\Delta) \\  0& +i\,n \cos(n\Delta) u(\Delta)
 \end{array} \right ]\,.
\end{eqnarray}

In order to assemble the results I consider first terms proportional
$\delta\left(k_{\rm x} - k'_{\rm x}\right)$ or $\delta(\Delta)$ and integrate 
immediately over $k_{\rm x}$, which contracts the integrand to 
$k_{\rm x} = k'_{\rm x}$. Equations (48), (49), and (55) together with (47)
give
\begin{eqnarray}
&& -\frac{4\pi G}{2\pi}\frac{\kappa}{-2B}\frac{\Sigma_0}{\kappa\sigma_u^2}
\Phi_{\bf{\rm  k'}} \Big[ 2 \pi\left(-2B\right)   
-2\pi\left(-2B\right)e^{-\Lambda\left(k_{\rm x} = k'_{\rm x}\right)} 
\nonumber \\ && \cdot \{
I_0\left(\Lambda\left(k_{\rm x} = k'_{\rm x}\right)\right)  
- 2 \sum_{n=1}^{\infty}\left(-1\right)^{\rm n}I_{\rm n}
\left(-\Lambda \left(k_{\rm x} = k'_{\rm x}\right)\right)\}\Big]\,.\nonumber \\
\end{eqnarray}
Since 
\begin{equation}
\sum_{n=1}^{\infty}\left(-1\right)^{\rm n} I_{\rm n} 
\left(-\Lambda\right) = \sum_{n=1}^{\infty} I_{\rm n}(\Lambda) = 
\frac{1}{2}\left(e^{\Lambda} - I_0\left(\Lambda\right)\right)
\end{equation}
(Abramowitz \& Stegun 1968, section 9.6) , the term (57) cancels out exactly. 
This leaves according to equations (47) and (56)
\begin{eqnarray}
&&-\frac{4\pi G}{2\pi}\frac{\kappa}{-2B}\frac{\Sigma_0}{\kappa\sigma_u^2}
\int^{+\infty}_{-\infty}dk_x \Phi_{\bf{k}}\nonumber\\ && \cdot
\exp -\frac{\sigma_u^2}{\kappa^2}\left[\frac{1}{2}
\left(k_{\rm x}^2 + {k'}_{\rm x}^2 \right) + \frac{\kappa^2}{4B^2}{k'}_{\rm y}^2
\right] 2\pi \frac{B}{A}\frac{\kappa}{|k'_{\rm y}|}\nonumber\\ && \cdot
\exp - \left(\frac{i\omega}{\kappa}\Delta\right) \\ && \cdot
2 \Big[
\sum^{\infty}_{n=1}(-n)\sin\left(n\Delta\right)\cos\left(n\left(\overline{w}
- \overline{w}'\right)\right)\left(-1\right)^{\rm n} 
I_{\rm n} \left(-\Lambda\right)
 \nonumber \\ && - n \cos \left(n\Delta\right)\sin\left(n\left(\overline{w}- 
\overline{w'}\right)\right)\left(-1\right)^{\rm n} I_{\rm n} 
\left(-\Lambda\right) \Big] u\left(\Delta\right) \,. \nonumber
\end{eqnarray}
The series 
\begin{equation}
\sum_{n=1}^{\infty}n\sin\left(n\left(\Delta + \overline{w} - 
\overline{w}'\right)\right)I_{\rm n}\left(\Lambda\right)
\end{equation}
can be simplified by using the identity
\begin{equation} 
n I_{\rm n}\left(\Lambda\right) = \frac{\Lambda}{2}
\left(I_{{\rm n}-1} \left(\Lambda\right) - I_{{\rm n}+1}
\left(\Lambda\right)\right)
\end{equation}
and rearranging it as
\begin{eqnarray} 
&&\frac{\Lambda}{2}I_0 \left(\Lambda\right)\sin\left(\Delta  + 
\overline{w} - \overline{w}'\right) \nonumber \\ && + 
\Lambda\sin \left(\Delta + 
\overline{w} - \overline{w}'\right)\sum_{n=1}^{\infty}I_{\rm n} 
\left(\Lambda\right)\cos\left(n\left(\Delta + \overline{w} - 
\overline{w}'\right)\right) \nonumber \\ &&
= \frac{\Lambda}{2}\sin
\left(\Delta + \overline{w} - \overline{w}'\right)\exp\left(
\Lambda\cos\left(\Delta + \overline{w} - \overline{w}'\right)\right)
\end{eqnarray}
(Abramowitz \& Stegun 1968, section 9.6). Furthermore, using the definitions of
$\Lambda$, $\overline{w}$, and $\overline{w}'$ I find the relations
\begin{eqnarray}
&&\Lambda\sin\left(\Delta + \overline{\omega} - \overline{\omega}'\right)
= \frac{\sigma_{\rm{u}}^{\rm{2}}}{\kappa^{\rm{2}}} \nonumber \\ &&
\cdot \{\frac{\kappa}{2B}k'_{\rm{y}}
\left(k'_{\rm{x}} - k_{\rm{x}}\right)\cos\Delta + \left(k_{\rm{x}}k'_{\rm{x}}
+ \frac{\kappa^{\rm{2}}}{4B^{\rm{2}}}{k'}_{\rm{y}}^{\rm{2}}\right)\sin\Delta\}
\end{eqnarray}
and
\begin{eqnarray}
&&\Lambda\cos\left(\Delta + \overline{\omega} - \overline{\omega}'\right)
= \frac{\sigma_{\rm{u}}^{\rm{2}}}{\kappa^{\rm{2}}} \\ &&
\cdot\{\left(k_{\rm{x}}k'_{\rm{x}}
 + \frac{\kappa^{\rm{2}}}{4B^{\rm{2}}}{k'}_{\rm{y}}^{\rm{2}}\right)\cos\Delta 
 - \frac{\kappa}{2B}k'_{\rm{y}}\left(k'_{\rm{x}} - k_{\rm{x}}\right)
 \sin\Delta\}\,,\nonumber
\end{eqnarray} 
so that the multiple integral (38) can be put into the final form
\begin{eqnarray}
&&-\frac{2\pi G\Sigma_{\rm{0}}}{A\kappa | k'_{\rm{y}}|}
\int_{-\infty}^{k'_{\rm{x}}} dk_{\rm x}\Phi_{\rm{\bf{k}}}                       
\{\left(k_{\rm{x}}k'_{\rm{x}} + \frac{\kappa^{\rm{2}}}{4B^{\rm{2}}}
{k'}_{\rm{y}}^{\rm{2}}\right)
\sin\frac{\kappa\left(k'_{\rm{x}} - 
k_{\rm{x}}\right)}{2Ak'_{\rm{y}}}\nonumber\\ && 
+ \frac{\kappa}{2B}k'_{\rm{y}}\left(k'_{\rm{x}} - k_{\rm{x}}\right)
\cos\frac{\kappa\left(k'_{\rm{x}} - k_{\rm{x}}\right)}{2Ak'_{\rm{y}}}\}
\exp - i\omega \frac{k'_{\rm{x}} - k_{\rm{x}}}{2Ak'_{\rm{y}}}\nonumber\\ &&
\cdot\exp - \frac{\sigma_{\rm{u}}^{\rm{2}}}{\kappa^{\rm{2}}}\Big[ \frac{1}{2}
(k_{\rm{x}}^{\rm{2}} + {k'}_{\rm{x}}^{\rm{2}}) + \frac{\kappa^{\rm{2}}}
{4B^{\rm{2}}}{k'}_{\rm{y}}^{\rm{2}}
- \left(k_{\rm{x}}k'_{\rm{x}} + \frac{\kappa^{\rm{2}}}{4B^{\rm{2}}}
{k'}_{\rm{y}}^{\rm{2}}\right)
\nonumber \\ && \cdot\cos\frac{\kappa\left(k'_{\rm{x}} - k_{\rm{x}}
\right)}{2Ak'_{\rm{y}}}
+ \frac{\kappa}{2B}k'_{\rm{y}}\left(k'_{\rm{x}} - k_{\rm{x}}\right)
\sin\frac{\kappa\left(k'_{\rm{x}} - k_{\rm{x}}\right)}{2Ak'_{\rm{y}}}\Big]\,.
\end{eqnarray}
In expression (65) positive wave numbers $k'_{\rm{y}}$ are assumed. In the case 
of negative wave numbers $k'_{\rm{y}}$ the integration limits are reversed 
to $\int_{k'_{\rm{x}}}^{+\infty}dk_{\rm{x}}$.

The fundamental integral equation is thus a Volterra equation,
\begin{equation}
\Phi_{\bf{\rm{k'}}} = \int_{-\infty}^{k'_{\rm{x}}} dk_{\rm{x}} {\cal K} 
\left(k_{\rm{x}},k'_{\rm{x}}\right)\Phi_{\bf{\rm{k}}}\,
\end{equation}
with the kernel
\begin{eqnarray}
&&{\cal K} \left(k_{\rm{x}} ,k'_{\rm{x}}\right) = \frac{\pi G \Sigma_{\rm{0}}}
{A\kappa |k'_{\rm{y}}|}\frac{1}{\sqrt{{k'}_{\rm{x}}^{\rm{2}} + 
{k'}_{\rm{y}}^{\rm{2}}}}\exp -i\omega\frac{k'_{\rm{x}} - k_{\rm{y}}}
{2Ak'_{\rm{y}}}\nonumber\\ && \cdot
\{\left(k_{\rm{x}} k'_{\rm{x}} + \frac{\kappa^{\rm{2}}}{4B^{\rm{2}}}
{k'}_{\rm{y}}^{\rm{2}}\right)\sin\frac{\kappa\left(k'_{\rm{x}} - 
k_{\rm{x}}\right)}{2Ak'_{\rm{y}}} \nonumber \\ &&
+ \frac{\kappa}{2B}k'_{\rm{y}}\left(
k'_{\rm{x}} - k_{\rm{x}}\right)\cos\frac{\kappa\left(k'_{\rm{x}} - 
k_{\rm{x}}\right)}{2Ak'_{\rm{y}}}\}\nonumber\\ && \cdot
\exp -\frac{\sigma_{\rm{u}}^{\rm{2}}}{\kappa^{\rm{2}}}\Big[\frac{1}{2}
\left(k_{\rm{x}}^{\rm{2}} + {k'}_{\rm{x}}^{\rm{2}}\right) + 
\frac{\kappa^{\rm{2}}}{4B^{\rm{2}}}{k'}_{\rm{y}}^{\rm{2}} - 
\left(k_{\rm{x}} k'_{\rm{x}} + \frac{\kappa^{\rm{2}}}{4B^{\rm{2}}}k'_{\rm{y}}
e^{\rm{2}}\right) \nonumber \\ && \cdot
\cos\frac{\kappa\left(k'_{\rm{x}} - k_{\rm{x}}\right)}{2Ak'_{\rm{y}}}
+ \frac{\kappa}{2B}k'_{\rm{y}}\left(k'_{\rm{x}} - k_{\rm{x}}\right)
\sin\frac{\kappa\left(k'_{\rm{x}} - k_{\rm{x}}\right)}{2Ak'_{\rm{y}}}\Big]\,.
\end{eqnarray}
The integral equation (66) is equivalent to that obtained by Julian \& 
Toomre (1966) although it is derived and presented here in a different way.

\section{Discussion of the Volterra equation}

Surprisingly the Volterra equation (66) has no solution. This can be seen by a
Fredholm discretization of the equation (Morse \& Feshbach 1953). The integral
equation is then replaced by a set of algebraic equations with a triangular 
coefficient matrix. The kernel (67) vanishes on the diagonal, $k'_{\rm{x}}
= k_{\rm{x}}$, so that the diagonal of the coefficient matrix is given by the
identity matrix from the left hand side of equation (66). The determinant of a
triangular matrix is equal to the product of the diagonal elements of the
matrix, which is equal to one in this case. This means that the set of
algebraic equations and thus the integral equation have no non--trivial
solutions. According to the theorem of Fredholm the inhomogenous Volterra 
equation
\begin{equation}
\Phi_{\bf{k'}} = \int_{-\infty}^{k'_{\rm{x}}}dk_{\rm{x}} {\cal K}
\left(k_{\rm{x}} , k'_{\rm{x}}\right)\Phi_{\rm{\bf{k}}} + r_{\rm{\bf{k'}}}
\end{equation}
has a unique solution, if the corresponding homogenous equation has no solution.
The solution can be expressed with the resolvent operator $\cal R$ as
\begin{equation}                                                                
\Phi_{\rm{\bf{k'}}} = r_{\rm{\bf{k'}}} + \int_{-\infty}^{k'_{\rm{x}}}dk_{\rm{x}}
{\cal R} \left(k_{\rm{x}} , k'_{\rm{x}}\right) r_{\rm{\bf{k}}}\,,               
\end{equation}
where the resolvent operator is given by a Neumann series,
\begin{equation}
{\cal R}\left(k_{\rm{x}} , k'_{\rm{x}}\right) = \sum_{n=1}^{\infty} 
{\cal K}^{\left({\rm n}\right)}\left(k_{\rm{x}} ,k'_{\rm{x}}\right)
\end{equation}
with 
\begin{eqnarray}
&&{\cal K}^{\left(1\right)}\left(k_{\rm{x}} , k'_{\rm{x}}\right) = 
{\cal K}\left(k_{\rm{x}} , k'_{\rm{x}}\right)\,,{\rm and} \nonumber \\
&&{\cal K}^{({\rm{n}})}\left(k_{\rm{x}} , k'_{\rm{x}}\right) = 
\int_{k_{\rm{x}}}^{k'_{\rm{x}}}dk''_{\rm{x}} 
{\cal K}\left(k''_{\rm{x}} , k'_{\rm{x}}\right) 
{\cal K}^{\rm{\left(n-1\right)}}\left(k_{\rm{x}} , k''_{\rm{x}}\right)\,.
\end{eqnarray} 
The inhomogeneity of equation (68) represents perturbations of the shearing
sheet,
which have not been included in the analysis so far. As will be discussed in 
the next section, these may be due to external forces or massive perturbations
within the disk, or any initial non--equilibrium states of the disk. 
Equation (69) describes then the response of the disk to these perturbations.
If they are absent, the shearing sheet does not develop any inhomogenous
spatial 
structures and perturbations of the velocity field. There is a technical point
concerning the functional dependence of the resolvent operator on the frequency 
$\omega$, which will become important later on. Equation (67) shows that the 
kernel ${\cal K} \left(k_{\rm{x}} , k'_{\rm{x}}\right)$ is proportional to
$\exp -i\omega\frac{k'_{\rm{x}} - k_{\rm{x}}}{2Ak'_{\rm{y}}}$. This 
proportionality is maintained under the iterations of equation (71), so that 
the resolvent operator has the same functional dependence as ${\cal K}$, 
${\cal R} \left(k_{\rm{x}} , k'_{\rm{x}}\right) \propto 
\exp -i\omega\frac{k'_{\rm{x}} - k_{\rm{x}}}{2Ak'_{\rm{y}}}$.

When deriving the Volterra equation (68), I have not considered the case
$k'_{\rm{y}} = 0$, which corresponds in disks with polar geometry to 
ring--like perturbations. It is straightforward to show using the formalism of
section (4) that in this case the disk response to a Fourier component of the
general potential perturbation is the corresponding Fourier component of the
general disk response, so that they form conjugate pairs for each wave number 
vector ${\bf k'}=\left(k'_{\rm{x}}, 0\right)$. When self--gravity is taken
into account, the result is a dispersion relation 
$\omega\left(k_{\rm{x}}\right)$, which is formally identical with that of 
Lin, Yuan, \& Shu (1969), although it is valid here only for
$k'_{\rm{y}} = 0$. Thus the case of $k'_{\rm{y}} = 0$ is not generic for cases 
of even small non--vanishing circumferential wave members. The same holds true
for Oort's constant $A$. The dynamical behaviour of a rigidly rotating shearing
sheet ($A=0$) or star disks, in general, is not generic for the behaviour of 
differentially rotating disks.                                                  

\section{Initial value problem}

If the shearing sheet is initially -- either at time minus infinity or at some
$t=t_{\rm{0}}$ -- not in equilibrium, this must be taken into account
explicitely in the Fourier transform of equation (20) with respect to the time
coordinate. The Laplace transform of the time derivative of the perturbation of
the distribution function gives in this case
\begin{eqnarray}
&&\int_{t_{\rm{0}}}^{\infty}dte^{-i\omega t}\frac{\partial\delta f}
{\partial t} = -\delta f \left(t_{\rm{0}}\right)e^{-i\omega t_0} + 
\int_{t_{\rm{0}}}^{\infty}dt\,i\,\omega e^{-i\omega t}\delta f \nonumber \\ &&=
-\delta f\left(t_{\rm{0}}\right)e^{\rm{-i\omega t_0}} + 
i\omega\delta f_{\rm{\omega}}\,.
\end{eqnarray}
This means a further inhomogenous term in equation (23). Equation (23) is
linear, and therefore the solution of the extended equation can be constructed
by superposing the solution of equation (23) in its original form as derived 
in section (4) and a solution of
\begin{eqnarray}
i\omega \delta f_{\rm{\omega}} + \kappa \frac{\partial\delta f_{\rm{\omega}}}
{\partial w_1} + \frac{A}{B}\left(J_{\rm{2}} - \Omega_{\rm{0}} r_{\rm{0}}\right)
\frac{\partial\delta f_{\rm{\omega}}}{\partial w_{\rm{2}}}\nonumber \\  - 
\delta f\left(t_{\rm{0}}\right)e^{-i\omega t_0} = 0\,.
\end{eqnarray}
In this way both inhomogenous terms of the extended equation (23) are taken 
into account. First, the inhomogenous term of equation (73) is Fourier 
transformed
\begin{eqnarray}
&&\delta f \left(t_{\rm{0}}\right) = \int_{-\infty}^{+\infty}dk_{\rm{x}}
\int_{-\infty}^{+\infty}dk_{\rm{y}}\delta f_{\rm{\bf{k}}}\left(t_{\rm{0}}\right)
\nonumber \\ && \cdot
\exp\Big[ik_{\rm{x}} \left(\frac{J_{\rm{2}} - \Omega_{\rm{0}} r_{\rm{0}}}{-2B} 
+ \sqrt{\frac{2J_{\rm{1}}}{\kappa}}\sin w_{\rm{1}}\right) \nonumber \\ &&
 + ik_{\rm{y}}
w_{\rm{2}} - ik_{\rm{y}} \frac{\sqrt{2\kappa J_{\rm{1}}}}{2B}\cos 
w_{\rm{1}}\Big]\,.
\end{eqnarray}
Again the background distribution function $f_{\rm{0}}$ is split off all terms 
in equation (73), in particular $\delta f_{\rm{\bf{k}}}\left(t_{\rm{0}}\right) =
f_{\rm{0}} f_{\rm{\bf{k}}}\left(t_{\rm{0}}\right)$, and a Fourier transform of 
equation (73) with respect to the angle variable $w_{\rm{2}}$ leads to
\begin{eqnarray}
&&i\omega f_{\rm{2,\omega}} + \kappa \frac{df_{\rm{2,\omega}}}{dw_{\rm{1}}} + 
ik_{\rm{y}} \frac{A}{B}\left(J_{\rm{2}} - \Omega_{\rm{0}} r_{\rm{0}}\right)
f_{\rm{2,\omega}}\nonumber\\&&
-\int_{-\infty}^{+\infty}dk_{\rm{x}} f_{\rm{\bf{k}}}\left(t_{\rm{0}}\right)
e^{-i\omega t_0}\nonumber \\ && \cdot
\exp ik_{\rm{x}} \Big[\frac{J_{\rm{2}} - \Omega_{\rm{0}}
r_{\rm{0}}}{-2B} + \sqrt{\frac{2J_{\rm{1}}}{\kappa}}\sin w_{\rm{1}} \Big]
\nonumber \\ && \cdot
\exp -ik_{\rm{y}} \frac{\sqrt{2\kappa J_{\rm{1}}}}{2B}\cos w_{\rm{1}} = 0\,.
\end{eqnarray}
The derivation of the solution of equation (75) proceeds in exactly 
the same way as with equation (23) in section (4), so that I report only the 
result using the notation from above
\begin{eqnarray}
&&f_{\rm{2,\omega}} = -\frac{1}{\kappa}
\int_{-\infty}^{+\infty}dk_{\rm{x}}
f_{\rm{\bf{k}}}\left(t_{\rm{0}}\right)e^{-i\omega t_0} \nonumber \\&& \cdot
\exp i\left(k_{\rm{x}}
\frac{J_{\rm{2}} - \Omega_{\rm{0}} r_{\rm{0}}}{-2B}\right) 
\frac{i}{2\sin\left(\pi\eta\right)} \nonumber \\ && \cdot
\int_{-\pi}^{\pi}d w'_{\rm{1}} \exp i \Big[\eta w'_{\rm{1}} 
- \xi\sin\left(w'_{\rm{1}} + w_{\rm{1}} - 
\overline{w}\right)\Big]\,. 
\end{eqnarray}
The disk response to the potential perturbation is now given by
\begin{equation}
\delta f_{\rm{\omega}} = f_{\rm{0}} \left(f_{\rm{1,\omega}} + f_{\rm{2,\omega}}
\right) \exp i\left(\omega t + k_{\rm{y}} w_{\rm{2}} \right)\,.
\end{equation}
The second part of the disk response has also to be considered in the Poisson 
equation. Again this can be done separately and the expression analogous to (38)
is given by
\begin{eqnarray}
&&-\frac{4\pi G}{\left(2\pi\right)^{\rm{2}}} \int_{0}^{2\pi}d w_{\rm{1}}
\int_{-\infty}^{+\infty}dw_{\rm{2}} \int_{0}^{\infty}dJ_{\rm{1}}
\int_{-\infty}^{+\infty}dJ_{\rm{2}} \frac{\kappa}{-2B}\nonumber \\ && \cdot
\frac{\Sigma_{\rm{0}}}{2\pi\sigma_{\rm{u}}^{\rm{2}}}\exp\left(-\frac{\kappa}
{\sigma_{\rm{u}}^{\rm{2}}}J_1\right) \nonumber \\ && \cdot
\int_{-\infty}^{+\infty}dk_{\rm{y}} \exp(ik_{\rm{y}} w_{\rm{2}})\frac{1}{\kappa}
\int_{-\infty}^{+\infty}dk_{\rm{x}} f_{\rm{\bf{k}}}\left(t_{\rm{0}} \right)
e^{-i\omega t_0} \nonumber \\ && \cdot
\exp \left(ik_{\rm{x}} \frac{J_{\rm{2}} - \Omega_{\rm{0}}
r_{\rm{0}}}{-2B}\right)\frac{i}{2\sin\left(\pi\eta\right)} \nonumber \\&&
\cdot\int_{-\pi}^{+\pi}d w'_{\rm{1}} \exp i\left[\eta w'_{\rm{1}} - \xi
\sin\left(w'_{\rm{1}} + w_{\rm{1}} - \overline{w}\right)\right] \nonumber \\ &&
\cdot\exp -i\Big[k'_{\rm{x}}\frac{J_{\rm{2}} - \Omega_{\rm{0}} r_{\rm{0}}}{-2B}
+k'_{\rm{x}} \sqrt{\frac{2J_{\rm{1}}}{\kappa}}\sin w_{\rm{1}} \Big]
 \nonumber \\&& \cdot \exp i \Big[
 k'_{\rm{y}} w_{\rm{2}} - k'_{\rm{y}} \frac{\sqrt{2\kappa J_{\rm{1}}}}{2B}
\cos w_{\rm{1}}\Big]\,.
\end{eqnarray}
The quadratures can be carried out in the same way as in section (6) and I
report again only the result,
\begin{eqnarray}
&&\frac{2\pi G \Sigma_{\rm{0}}}{A|k'_{\rm{y}}|} \int_{-\infty}^{k'_x}dk_{\rm{x}}
f_{\rm{k_x ,k'_y}}\left(t_{\rm{0}} \right)e^{-i\omega t_0}
\exp -i \omega \frac{k'_{\rm{x}} - k_{\rm{x}}}{2Ak'_y} \nonumber \\
&&\cdot\exp -\frac{\sigma_{\rm{u}}^{\rm{2}}}{\kappa^{\rm{2}}}\Big[\frac{1}{2}
\left(k_{\rm{x}}^{\rm{2}} + \frac{\kappa^{\rm{2}}}{4B^{\rm{2}}}{k'}_{\rm{y}}
^{\rm{2}} \right) + \frac{1}{2}\left({k'}_{\rm{x}}^{\rm{2}} + 
\frac{\kappa^{\rm{2}}}{4B^{\rm{2}}}{k'}_{\rm{y}}^{\rm{2}} \right) \nonumber \\
&&- \left(k'_{\rm{x}} k_{\rm{x}} + \frac{\kappa^{\rm{2}}}{4B^{\rm{2}}}
{k'}_{\rm{y}}^{\rm{2}} \right) \cos\frac{\kappa\left(k'_{\rm{x}} - k_{\rm{x}}
\right)}{2Ak'_y} \nonumber \\ &&
+ \frac{\kappa}{2B}k'_{\rm{y}}\left(k'_{\rm{x}} - k_{\rm{x}} \right)
\sin\frac{\kappa\left( k'_{\rm{x}} - k_{\rm{x}} \right)}{2Ak'_{\rm{y}}}\Big]\,,
\end{eqnarray}
where positive wave numbers $k'_{\rm y}$ are assumed. In the case of negative
wave numbers $k'_{\rm y}$ the integration limits have to be reversed to
$\int_{k'_{\rm x}}^{\infty} dk_{\rm x}$.
Thus the inhomogenous term of the integral equation (68) can be written as
\begin{equation}
r_{\rm{\bf{k'}}} = \int_{-\infty}^{k'_{\rm x}}dk_{\rm{x}} {\cal L} 
\left(k_{\rm{x}} ,k'_{\rm{x}}\right) f_{\rm{k_{\rm x} , k'_{\rm x}}}
\left(t_{\rm{0}}\right)e^{-i\omega t_0}
\end{equation}
with
\begin{eqnarray}
&&{\cal L}\left(k_{\rm{x}} , k'_{\rm{x}}\right) = - \frac{\pi G\Sigma_{\rm{0}}}
{A|k'_{\rm{y}}|}\frac{1}{\sqrt{{k'}_{\rm{x}}^{\rm{2}} + 
{k'}_{\rm{y}}^{\rm{2}}}}\exp -i\omega\frac{k'_{\rm{x}} - k_{\rm{x}}}
{2Ak'_{\rm{y}}} \nonumber\\ && \cdot
\exp -\frac{\sigma_{\rm{u}}^{\rm{2}}}{\kappa^{\rm{2}}}\Big[\frac{1}{2} 
\left(k_{\rm{x}}^{\rm{2}} + \frac{\kappa^{\rm{2}}}{4B^{\rm{2}}}
{k'}_{\rm{y}}^{\rm{2}} \right) + \frac{1}{2}\left({k'}_{\rm{x}}^{\rm{2}}
+ \frac{\kappa^{\rm{2}}}{4B^{\rm{2}}}{k'}_{\rm{y}}^{\rm{2}} \right) \nonumber \\
&&- \left(k'_{\rm{x}} k_{\rm{x}} + \frac{\kappa^{\rm{2}}}{4B^{\rm{2}}}
{k'}_{\rm{y}}^{\rm{2}} \right) \cos\frac{\kappa\left(k'_{\rm{x}} - k_{\rm{x}} 
\right)}{2Ak'_{\rm{y}}} \nonumber \\
&&+ \frac{\kappa}{2B}k'_{\rm{y}}\left(k'_{\rm{x}} - k_{\rm{x}} \right)
\sin\frac{\kappa\left( k'_{\rm{x}} - k_{\rm{x}} \right)}{2Ak'_{\rm{y}}}\Big]\,,
\end{eqnarray}
where the $\omega$-dependent term has exactly the same form as in the kernel 
function ${\cal K}$.
The general solution of the integral equation (68) has then the specific form
\begin{eqnarray}
&&\Phi_{\rm{\bf{k'}}} = \int_{-\infty}^{k'_{\rm x}}dk_{\rm{x}} {\cal L} 
\left(k_{\rm{x}} ,k'_{\rm{x}}\right) f_{\rm{k_{\rm x} ,k'_{\rm x}}}
\left(t_{\rm{0}} \right)e^{-i\omega t_0}  \\ &&
+ \int_{-\infty}^{k'_{\rm x}}dk_{\rm{x}} {\cal R} 
\left(k_{\rm{x}} , k'_{\rm{x}} \right) \int_{-\infty}^{k_{\rm x}}dk''_{\rm{x}}
{\cal L} \left(k''_{\rm{x}} , k_{\rm{x}} \right) f_{\rm{k''_{\rm x} , 
k'_{\rm y}}}\left(t_{\rm{0}} \right) e^{-i\omega t_0}\,.\nonumber
\end{eqnarray}
The potential (82) is still defined in Fourier space, but, when transformed
back to time and spatial coordinates, gives the description of the evolution of
the disturbance of the shearing sheet. Illustrative examples are described in
the next section.

It is instructive to consider the behaviour of the solution (82) 
in the limit $t\to\infty$ $\left(k'_{\rm{y}} > 0\right)$,
\begin{eqnarray}
&&\lim_{t \to \infty} \frac{1}{2\pi}\int_{-\infty}^{+\infty}
d\omega e^{i\omega t} \Phi_{\rm{\bf{k'}}, \omega} \\ &&
= \lim_{t \to \infty} \int_{-\infty}^{k'_{\rm x}}dk_{\rm{x}}
f_{\rm{k_{\rm x} , k'_{\rm y}}}(t_0) \nonumber \\&& \cdot
 2Ak'_{\rm{y}} \delta\left(k_{\rm{x}} - k'_{\rm{x}} + 2Ak'_{\rm{y}}\left(t - 
 t_{\rm{0}} \right)\right) \tilde{\cal L} \left(k_{\rm{x}} , k'_{\rm{x}}\right)
\nonumber \\ &&
+\lim_{t \to \infty} \int_{-\infty}^{k'_{\rm x}}dk_{\rm{x}}
 \tilde{\cal R} \left(k_{\rm{x}} , k'_{\rm{x}}\right) \int_{-\infty}^{k_{\rm x}}
 dk''_{\rm{x}} \tilde{\cal L} \left(k''_{\rm{x}} , k_{\rm{x}}\right)
  f_{\rm{k''_{\rm x} , k'_{\rm y}}} \left(t_0\right)
   \nonumber \\ && \cdot
2Ak'_{\rm{y}} \delta\left(k''_{\rm{x}} - k'_{\rm{x}} + 2Ak'_{\rm{y}}
\left(t - t_{\rm{0}} \right)\right) \nonumber \\ &&
= \lim_{t \to \infty} 2Ak'_{\rm{y}} \tilde{\cal L} \left(k'_{\rm{x}} - 
2Ak'_{\rm{y}}\left(t - t_{\rm{0}} \right),k'_{\rm{x}}\right) \nonumber \\&&
\cdot f_{\rm{k'_{\rm x}-2Ak'_{\rm y}(t-t_0),k'_{\rm y}}}\left(t_{\rm{0}}\right)
 \nonumber \\ &&
+ \lim_{t \to \infty} 2Ak'_{\rm{y}} \int_{-\infty}^{k'_{\rm x}}dk_{\rm{x}} 
\tilde{\cal R} \left(k_{\rm{x}} , k'_{\rm{x}} \right) \tilde{\cal L}
\left(k'_{\rm{x}} - 2Ak'_{\rm{y}} \left( t - 
t_{\rm{0}} \right), k_{\rm{x}}\right) \nonumber \\ && \cdot
f_{\rm{k'_{\rm x}-2Ak'_{\rm y}(t-t_0),k'_{\rm y}}}\left(t_{\rm{0}}\right)\,,
\nonumber
\end{eqnarray}
where operators with an overhead tilde denote again operators with the
$\omega$--dependent terms (cf. section 7 and equation 81) split off. As 
can be seen from equation (81) the $\tilde{\cal L}$ functions decay as
\begin{eqnarray}
&&\lim_{t \to \infty} \tilde{\cal L} \left(k'_{\rm{x}} - 2Ak'_{\rm{y}} 
\left(t - t_{\rm{0}} \right), k'_{\rm{x}}\right) \nonumber \\ 
&& = \lim_{t \to \infty} \tilde{\cal L} \left(k'_{\rm{x}} - 2Ak'_{\rm{y}} 
\left(t - t_{\rm{0}} \right), k_{\rm{x}}\right) \nonumber \\ &&
\propto \lim_{t \to \infty} \exp -\frac{\sigma_{\rm{u}}^{\rm{2}}}
{2\kappa^{\rm{2}}}\left(2Ak'_{\rm{y}} \left(t-t_{\rm{0}} \right)\right)^{\rm{2}}
 = 0\,,
\end{eqnarray}
which means that the perturbation will die out eventually. This was first noted
by Julian \& Toomre (1966). The physical reason is that the effective radial
wave length of the perturbation, $\pi / \left(Ak'_{\rm{y}}\left(t - t_{\rm{0}}
\right)\right)$, becomes much smaller than the typical epicycle size of the
stellar orbits, measured by $\sigma_{\rm{u}} / \kappa$, so that the disk can no 
longer sustain such a perturbation. This is not the case for gas disks 
(Goldreich \& Tremaine 1978).

\section{Illustrative examples}

The characteristics of the solutions of the Volterra integral equation (68)
become immediately clear, when the shearing sheet is assumed to be perturbed 
initially by a single sinusoidal wave with a wave number 
${\bf{k}}^{\rm{in}} \left(k_{\rm{y}}^{\rm{in}} > 0\right)$,
\begin{equation}
\delta f_{\rm{\bf{k}}} \left(t_{\rm{0}} = 0\right) = 
f_{\rm{k_{\rm y}^{\rm in}}}^{\rm{in}}\delta\left(k_{\rm{x}} - 
k_{\rm{x}}^{\rm{in}}\right)\,.
\end{equation}
According to equation (80) this implies an inhomogeneity of the Volterra
equation
\begin{equation}
r_{\rm{k'_{\rm x} , k_{\rm y}^{\rm in}}} =
f_{\rm{k_{\rm y}^{\rm in}}}^{\rm{in}} {\cal L}\left(k_{\rm{x}}^{\rm{in}} , 
k'_{\rm{x}}\right)\,,
\end{equation}
and the solution of the Volterra equation is given according to equation (82)
by
\begin{eqnarray}
&&\Phi_{\rm{k'_{\rm x} ,k_{\rm y}^{\rm in}}} =
f_{\rm{k_{\rm y}^{\rm in}}}^{\rm{in}} {\cal L} \left(
k_{\rm{x}}^{\rm{in}},k'_{\rm{x}}\right) \nonumber \\ &&+ 
\int_{k_{\rm x}^{\rm in}}^{k'_{\rm x}}dk_{\rm{x}} {\cal R} \left(k_{\rm{x}} ,
k'_{\rm{x}} \right) f_{\rm{k_{\rm y}^{\rm in}}}^{\rm{in}} {\cal L} 
\left(k_{\rm{x}}^{\rm{in}},k_{\rm{x}} \right)\,.
\end{eqnarray}
Transforming this back to time and spatial coordinates leads to
\begin{eqnarray}
&&\Phi_{\rm{k'_{\rm x} , k_{\rm y}^{\rm in}, t}} = \frac{1}{2\pi}
\int_{-\infty}^{+\infty}d\omega e^{i\omega t} \Phi_{\rm{k'_{\rm x} , 
k_{\rm y}^{\rm in}, \omega}} \nonumber \\ &&
= f_{\rm{k_{\rm y}^{\rm in}}}^{\rm{in}} \tilde{\cal L} 
\left(k_{\rm{x}}^{\rm{in}},k'_{\rm{x}} \right) \delta \left(t + 
\frac{k_{\rm{x}}^{\rm{in}} - k'_{\rm{x}}}{2A{k}_{\rm{y}}^{\rm{in}}} \right)
 \\ &&
+ f_{\rm{k_{\rm y}^{\rm in}}}^{\rm{in}} \int_{k_{\rm x}^{\rm in}}^{k'_{\rm x}}
dk_{\rm{x}} \tilde{\cal R} \left(k_{\rm{x}} ,k'_{\rm{x}}\right) \tilde{\cal L}
\left(k_{\rm{x}}^{\rm{in}} , k_{\rm{x}} \right)\delta\left(t +
\frac{k_{\rm{x}}^{\rm{in}} - k'_{\rm{x}}}{2Ak_{\rm{y}}^{in}}\right) \nonumber
\end{eqnarray}
and then 
\begin{eqnarray}
&&\delta\Phi\left(x,y,t\right) = \frac{1}{2\pi}\int_{-\infty}^{+\infty}
dk'_{\rm{x}} e^{i\left[k'_{\rm x} x + k_{\rm y}^{\rm in} y \right]}
\Phi_{\rm{k'_{\rm x} , k_{\rm y}^{\rm in}}}\left(t\right) \nonumber\\&&
= \frac{Ak_{\rm{y}}^{\rm in}}{\pi}f_{\rm{k_{\rm y}^{\rm in}}}^{\rm{in}}
\{\tilde{\cal L }\left(k_{\rm{x}}^{\rm{in}}, k_{\rm{x}}^{\rm{in}} + 
2Ak_{\rm{y}}^{\rm{in}}t\right) \nonumber \\ &&
+ \int_{k_{\rm{x}}^{\rm{in}}}^{k_{\rm x}^{\rm in}+2Ak_{\rm y}^{\rm in}t}
dk_{\rm{x}} \tilde{\cal R}
\left(k_{\rm{x}} , k_{\rm{x}}^{\rm{in}} + 2Ak_{\rm{y}}^{\rm{in}}t\right) 
\tilde{\cal L} \left(k_{\rm{y}}^{\rm{in}}, k_{\rm{x}}\right)\} \nonumber \\&&
\cdot\exp i\left[\left(k_{\rm{x}}^{\rm{in}} + 2Ak_{\rm{y}}^{\rm{in}}t\right)x + 
k_{\rm{y}}^{\rm{in}}y\right]\,,
\end{eqnarray}
where the overhead tildes mean again that the $\omega$-dependent terms have 
been split off the operators $\cal L$ and $\cal R$ (cf.~the remarks at the end 
of section 7). An analogous expression can be derived for the surface density
of the perturbation $\delta\Sigma$. Inserting (cf.~equation 34),
\begin{equation}
\Sigma_{\rm{\bf{k}}} = - \frac{\sqrt{k_{\rm{x}}^{\rm{2}} + 
k_{\rm{y}}^{\rm{2}}}}{2\pi G}\Phi_{\rm{\bf{k}}}\,,
\end{equation}
into equation (89) gives immediately
\begin{eqnarray}
&&\delta\Sigma\left(x,y,t\right) = -\frac{1}{2\pi G}\frac{Ak_{\rm{y}}^{\rm in}}
{\pi} f_{\rm{k_{\rm y}^{\rm in}}}^{\rm{in}}\sqrt{\left(k_{\rm{x}}^{\rm{in}} +
2Ak_{\rm{y}}^{\rm in} t\right)^{\rm{2}} + {k_{\rm{y}}^{\rm{in}}}^{\rm{2}}} 
\nonumber \\&&
\cdot\{\tilde{\cal L} \left(k_{\rm{x}}^{\rm{in}}, k_{\rm{x}}^{\rm{in}} + 
2Ak_{\rm{y}}^{\rm{in}} t\right) + \int_{k_{\rm x}^{\rm in}}^{k_{\rm x}^{\rm in}
+2Ak_{\rm y}^{\rm in}t} 
dk_{\rm x} \nonumber \\ && \cdot
\tilde{\cal R} \left(k_{\rm{x}}^{\rm{in}}, k_{\rm{x}}^{\rm{in}} + 
2Ak_{\rm{y}}^{\rm{in}} t\right) \tilde{\cal L} \left(k_{\rm{x}}^{\rm{in}}, 
k_{\rm{x}}\right)\} \nonumber \\ &&
\cdot \exp i\left[\left(k_{\rm{x}}^{\rm{in}} + 2Ak_{\rm{y}}^{\rm{in}} 
t\right)x + k_{\rm{y}}^{\rm{in}}y\right]\,.
\end{eqnarray}
The spatial pattern described by equation (91) is still a sinusoidal wave, 
however with shearing wave crests. The wave crests, which are oriented
perpendicular to the wave number vector, are defined by lines
\begin{equation}
\left(k_{\rm{x}}^{\rm{in}} + 2Ak_{\rm{y}}^{\rm{in}}t\right)x + 
k_{\rm{y}}^{\rm{in}}y = const. \,,
\end{equation}
which tilt at the rate 
\begin{equation}
\frac{d}{dt}\left(\frac{\Delta y}{\Delta x}\right) = 2A\,.
\end{equation}
This corresponds exactly to the linear shear flow (11) of the shearing sheet.
The amplitude function in equation (91) describes the growth and damping of the
perturbation. As will be illustrated below, the amplitude function shows also
some oscillatory behaviour, so that the solutions of the Volterra equation 
have also to some extent the characteristics of moving wave trains. For this 
complex phenomenon Toomre (1981) has coined the term `swing-amplification'.

Unfortunately the amplitude function in equation (91) cannot be calculated 
analytically. However it needs only very modest numerical work to solve the 
original Volterra equation by Fredholm discretization numerically. The 
coefficient matrix of the resulting set of algebraic equations is already in
triangular form and extremely well behaved, because the kernel function (67) 
falls off in both the $k_{\rm x}$ and $k'_{\rm x}$ directions, respectively, 
tapered by gaussians. The Volterra equation can be written in dimensionless
form, if the wave numbers are measured in units of the critical wave number,
\begin{equation}
k_{\rm{crit}} = \frac{\kappa^{\rm{2}}}{2\pi G\Sigma_{\rm{0}}}\,,
\end{equation}
and the radial velocity dispersion $\sigma_{\rm{u}}$ is expressed in terms of 
the Toomre stability parameter $Q$
\begin{equation}
Q = \frac{\kappa \sigma_{\rm{u}}}{3.36G\Sigma_{\rm{0}}}\,.
\end{equation}
The numerical coefficient in the argument of the exponential function of the
 kernel (67) is then given by
\begin{equation}
\frac{\sigma_{\rm{u}}^{\rm{2}}}{\kappa^{\rm{2}}}k_{\rm{crit}}^{\rm{2}}
 = 0.2856 Q^{\rm{2}}
\end{equation}
Thus the kernel $\cal K$ and the $\cal L$ function are parameterized by the
values of $A/\Omega_{\rm{0}}$ and $Q$. The numerical scheme to solve the
integral equation is briefly described in the appendix.

In Fig.~1 the amplitude function of equation (91)
is shown for the case of $A/\Omega_{\rm{0}} = 0.5$ and $Q = 1.4$. The
amplification factors are drawn as function of $k_{\rm{x}}^{\rm{eff}} = 
k_{\rm{x}}^{\rm{in}} + 2Ak_{\rm{y}}^{\rm{in}}t$ abscissae for increasing 
circumferential wave numbers $k_{\rm{y}}^{\rm{in}}$. The sinusoidal waves have 
been incited with unit amplitude and with an initial radial wave number 
$k_{\rm{x}}^{\rm{in}} = -2\,k_{\rm{crit}}$. As can be seen from Fig.~1 maximum 
amplification is achieved, if the circumferential wave number is about
$0.5\,k_{\rm{crit}}$. This is the exact equivalent of the `X=2' criterion of
Toomre (1981). The maximum of amplification is reached, when the effective
radial wave number is $k_{\rm{x}}^{\rm{eff}} \approx\,2 k_{\rm{crit}}$, which 
corresponds to fairly open trailing `spiral' arms. The duration of the
amplification event is given by
\begin{equation}
t = \frac{k_{\rm{x}}^{\rm{eff}} - k_{\rm{x}}^{\rm{in}}}
{2Ak_{\rm{x}}^{\rm{in}}}\,,
\end{equation}
and is thus of the order of an orbital period $2\pi / \Omega_{\rm{0}}$. It can
be also seen from Fig.~1 that, if the circumferential wave number
$k_{\rm{y}}^{\rm{in}}$ becomes very small, which corresponds to tightly wound
`spiral' arms, the duration of amplification is fairly long and approaches the 
asymptotic WKB limit (Lin, Yuan \& Shu 1969).

\begin{figure} [h]
\begin{center}
\epsfxsize=8.7cm
   \leavevmode
     \epsffile{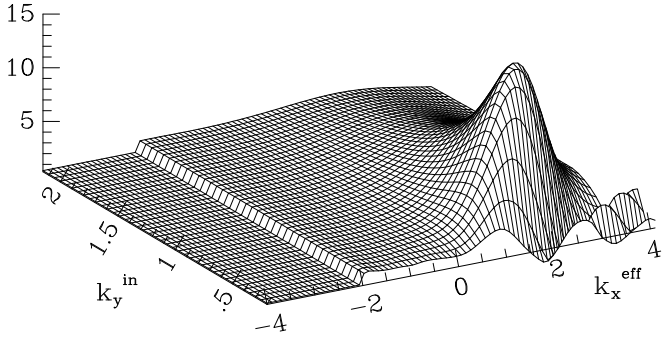}
\caption{Amplification of shearing density waves. $k_{\rm y}^{\rm in}$ denotes
the -- constant -- circumferential wave numbers in units of $k_{\rm crit}$.
$k_{\rm x}^{\rm eff}$ denotes the effective radial wave number and is given by
$k_{\rm x}^{\rm in} +2 A k_{\rm y}^{\rm in} t$, so that it increases from 
left to right during a swing--amplification event. The initial radial wave
number is chosen as $k_{\rm x}^{\rm in} = -2 k_{\rm crit}$, which corresponds 
to initially leading arms. The parameters of the disk model are $A/\Omega_0$ =
0.5 and $Q$ = 1.4.}
\label{fig1}
   \end{center}
   \end{figure}

While in Fig.~1 the arms were initially leading, in Fig.~2 the amplification of
initially trailing arms is shown. As can be seen from Fig.~2 there is hardly 
any amplification, which illustrates the well known result (Toomre 1981) that
the swing--amplification mechanism requires leading input in order to amplify 
density waves.

\begin{figure} [h]
\begin{center}
\epsfxsize=8.7cm
   \leavevmode
     \epsffile{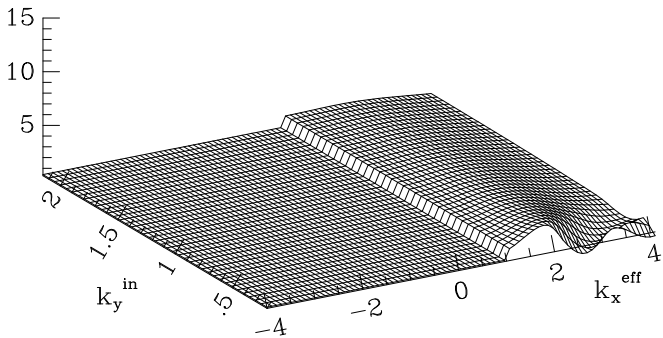}
\caption{Same as in Fig.~1, but the initial radial wave number
is chosen as $k_{\rm x}^{\rm in} = + 1 k_{\rm crit}$. The parameters of the 
disk model are again $A/\Omega_0$ = 0.5 and $Q$ = 1.4}
\label{fig2}
   \end{center}
   \end{figure}

\section{Parameter study}

The solutions of the Volterra equation are parametrized by the values of 
$A/\Omega_{\rm{0}}$ and $Q$. The peak of maximum amplification in wave number 
space shifts, when Oort's constant $A$ is varied. I have calculated a grid of
models with varying values of $A/\Omega_{\rm{0}}$ keeping the Toomre parameter
constant at $Q = 1.4$ and have determined for each model the circumferential
wave numbers for which peak amplification is reached. These can be fitted 
empirically by a relation of the form
\begin{equation}
\frac{k_{\rm{y,max}}}{k_{\rm{crit}}} = 1.932 - 5.186\left(\frac{A}
{\Omega_{\rm{0}}}\right) + 4.704\left(\frac{A}{\Omega_{\rm{0}}}
\right)^{\rm{2}}
\end{equation}
over the range $0.1 \leq \left(\frac{A}{\Omega_{\rm{0}}}\right) \leq 0.5$,
which corresponds to rising rotation curves. This domain is particularly
relevant, because in spiral galaxies the optically 
visible star disks often extend only over regions, where the rotation curves are
still rising. Relation (98) is in perfect agreement with the result of          
Athanassoula (1984), who sudied the influence of the shape of the rotation 
curve on swing--amplification by a different approach.

\begin{figure} [h]
\begin{center}
\epsfxsize=8.7cm
   \leavevmode
     \epsffile{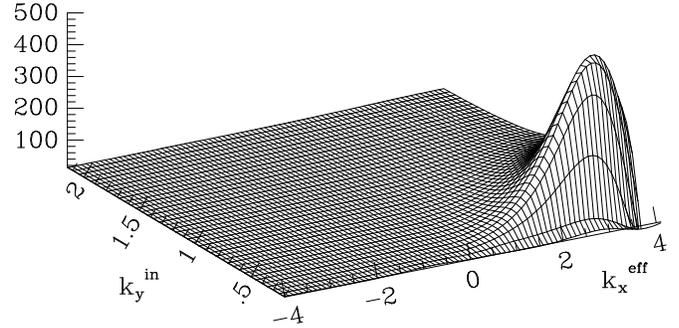}
\caption{Same as in Fig.~1, but the parameters of the disk model have been 
changed to $A/\Omega_0$ = 0.5 and $Q$ = 1.}
\label{fig3}
   \end{center}
   \end{figure}
   
The amplification factor itself depends strongly on the value of the Toomre 
stability parameter. In Figs.~3 and 4 amplitude functions are shown which
correspond to $Q = 1$ and $Q = 2$, respectively. As can be seen from Fig.~3
amplification becomes very high, when $Q$ approaches the limit of 
Jeans instability $Q = 1$, while in the case $Q = 2$ the disk is dynamically 
too hot to develop any structure.

\begin{figure} [h]
\begin{center}
\epsfxsize=8.7cm
   \leavevmode
     \epsffile{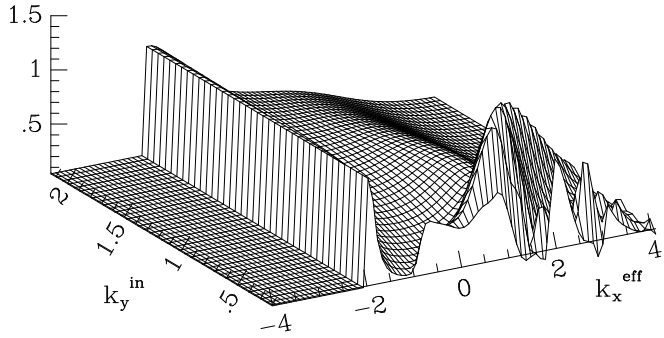}
\caption{Same as in Fig.~1, but the parameters of the disk model have been 
changed to $A/\Omega_0$ = 0.5 and $Q$ = 2.}
\label{fig4}
   \end{center}
   \end{figure}

\section{External perturbations}

When the shearing sheet is subjected to an external potential perturbation
$\Phi^{\rm{ext}}$, the sheet will respond to it and the response generates its
own gravitational potential $\Phi^{\rm{int}}$. The overall gravitational 
potential is obviously given by
\begin{equation}
\Phi = \Phi^{\rm{int}} + \Phi^{\rm{ext}} ,
\end{equation}
This has to be inserted into the Boltzmann equation, which leads to the 
determination of the distribution function of the perturbed sheet and 
eventually to the right hand side of equation (35). Into the left hand side of 
equation (35), however, only the gravitational potential generated by the 
perturbed sheet, $\Phi^{\rm{int}} = \Phi - \Phi^{\rm{ext}}$, has to be inserted.
If the external potential perturbation is Fourier transformed with respect to
time and spatial coordinates, its contribution to the left hand side of equation
(35) can be calculated using equation (36) leading to an inhomogeneity of the
Volterra equation (68)
\begin{equation}
r_{\rm{\bf{k'}}} = \Phi^{\rm{ext}}_{\rm{\bf{k'}},\omega}.
\end{equation}

\subsection{Excitation of a density wave}

If the shearing sheet is perturbated by an impulse of the form
\begin{equation}
\Phi^{\rm{ext}} = \Phi_{\rm{0}} \exp \{ i\left[k_{\rm{x}}^{\rm{in}}x +
k_{\rm{y}}^{\rm{in}}y\right] - \frac{1}{2}\left(\frac{t}{T}\right)^{\rm{2}}\}\,,
\end{equation}
where the time constant T indicates the length of its duration, the sheet
responds with a swing amplified density wave. The Fourier transform of equation 
(101) with respect to time and spatial coordinates leads to an inhomogeneity of
the Volterra equation (68) of the form
\begin{equation}
r_{\rm{\bf{k'}}} = \left(2\pi\right)^{\rm{\frac{5}{2}}}\Phi_{\rm{0}} T 
\delta \left(k'_{\rm{x}} - k_{\rm{x}}^{\rm{in}}\right)\delta\left(k'_{\rm{y}}
- k_{\rm{y}}^{\rm{in}}\right)e^{\rm{-\frac{1}{2}\omega^2 T^2}}\,.
\end{equation}
The solution of the Volterra equation is given according to equation (69) by
\begin{eqnarray}
&&\Phi_{\rm{\bf{k'}}} = \left(2\pi\right)^{\rm{\frac{5}{2}}}\Phi_{\rm{0}} T 
\delta \left(k'_{\rm{x}} - k_{\rm{x}}^{\rm{in}}\right)\delta\left(k'_{\rm{y}}
- k_{\rm{y}}^{\rm{in}}\right)e^{\rm{-\frac{1}{2}\omega^2 T^2}}\nonumber\\ &&
+ \int_{-\infty}^{k'_{\rm x}}dk_{\rm{x}} {\cal R} \left(k_{\rm{x}} ,
k'_{\rm{x}} \right)\left(2\pi\right)^{\rm{\frac{5}{2}}}\Phi_{\rm{0}} T \delta 
\left(k_{\rm{x}} - k_{\rm{x}}^{\rm{in}}\right)\delta\left(k'_{\rm{y}}
- k_{\rm{y}}^{\rm{in}}\right)\nonumber \\ &&
\cdot e^{\rm{-\frac{1}{2}\omega^2 T^2}}\,.
\end{eqnarray}
Transforming this back to time and spatial coordinates leads to
\begin{eqnarray}
&&\Phi\left(x,y,t\right) = \Phi_{\rm{0}} \exp \{i\left[k_{\rm{x}}^{\rm{in}}x + 
k_{\rm{y}}^{\rm{in}}y\right] - \frac{1}{2}\left(\frac{t}{T}\right)^{\rm{2}}\}
 \nonumber\\ &&
+ \frac{1}{2\pi}\int_{-\infty}^{+\infty}dk'_{\rm{x}}e^{i\left[k'_{\rm x} x
+ k_{\rm y}^{\rm in}y\right]}\int_{-\infty}^{k'_{\rm x}}dk_{\rm{x}} 
\tilde{\cal R} \left(k_{\rm{x}} ,k'_{\rm{x}}\right)\nonumber \\ &&
\cdot \frac{1}{2\pi} \int_{-\infty}^{+\infty}d\omega e^{i\omega (t
- \frac{k'_{\rm x} - k_{\rm x}}{2Ak_{\rm y}^{\rm in}})} \cdot
\left(2\pi\right)^{\rm{\frac{5}{2}}}\Phi_{\rm{0}} T \delta\left(k_{\rm{x}} 
- k_{\rm{x}}^{\rm{in}}\right)
 \nonumber \\ && \cdot e^{\rm{-\frac{1}{2}\omega ^2T^2}}\,,
\end{eqnarray}
where the $\omega$-dependent term of the resolvent operator $\cal R$ has been
written separately (cf. section 7). Carrying out the integrations gives
\begin{eqnarray}
&&\Phi\left(x,y,t\right) = \Phi_{\rm{0}} \exp \{i\left[k_{\rm{x}}^{\rm{in}}x +
k_{\rm{y}}^{\rm{in}}y\right] - \frac{1}{2}\left(\frac{t}{T}\right)^{\rm{2}}\}
 \nonumber\\ &&
+ \Phi_{\rm{0}} \int_{k_x^{\rm in}}^{+\infty}dk'_{\rm{x}} e^{i
\left[k'_{\rm x} x + k_{\rm y}^{\rm in}y\right]}\tilde{\cal R}
\left(k_{\rm{x}}^{\rm{in}},k'_{\rm{x}} \right)\nonumber\\ &&
\cdot \exp -\frac{1}{2T^{\rm{2}}}\left(t + \frac{k_{\rm{x}}^{\rm{in}}
- k'_{\rm{x}}}{2Ak_{\rm{y}}^{\rm{in}}}\right)^{\rm{2}}\,.
\end{eqnarray}
If the time constant is small, equation (105) can be approximately written as
\begin{eqnarray}
&&\Phi\left(x,y,t\right) \approx \Phi_{\rm{0}} \exp \{i\left[k_{\rm{x}}^{\rm{in}
}x + k_{\rm{y}}^{\rm{in}}y\right] -                                             
\frac{1}{2}\left(\frac{t}{T}\right)^{\rm{2}}\} \nonumber\\ &&
+ \widehat{\Phi_{\rm{0}}} \exp i\left[\left(k_{\rm{x}}^{\rm{in}}+2A
k_{\rm{y}}^{\rm{in}}
t\right)x + k_{\rm{y}}^{\rm{in}}y\right]\nonumber \\ && \cdot \tilde{\cal R}
\left(k_{\rm{x}}^{\rm{in}},k_{\rm{x}}^{\rm{in}} + 2Ak_{\rm{y}}^{\rm{in}}t\right)
\,.
\end{eqnarray}
The first term in equation (105) represents obviously the initial impulse, 
whereas the second term describes a swing amplified density wave exactly of the
kind already found in the treatment of the initial value problem in section (9)
(cf.~equation 89).

\subsection{Response to a point mass}                                           

The gravitational potential of a point mass resting at the origin of the
shearing sheet is given by
\begin{equation}
\Phi^{\rm{ext}} = -\frac{Gm}{r}\Big|_{\rm{z=0}} = -\frac{Gm}{\sqrt{x^{\rm{2}}
 + y^{\rm{2}}}}\,.
\end{equation}
Its Fourier transform with respect to spatial coordinates can be calculated 
using formulae (3.754) and (6.671) of Gradshteyn \& Ryzhik (1965) giving
\begin{equation}
\Phi_{\rm{\bf{k'}}}^{\rm{ext}} = -\frac{2\pi Gm}{\sqrt{{k'}_{\rm{x}}^{\rm{2}} 
+{k'}_{\rm{y}}^{\rm{2}}}}\,,
\end{equation}
so that the inhomogeneity of the Volterra equation has the form
\begin{equation}
r_{\rm{\bf{k'}}} = -\frac{4\pi^2 Gm}{\sqrt{{k'}_{\rm{x}}^{\rm{2}} + 
{k'}_{\rm{y}}^{\rm{2}}}}\delta(\omega)\,. 
\end{equation}
According to equation (69) the gravitational of the potential of the
perturbation is then
\begin{eqnarray}
&&\Phi_{\rm{\bf{k'}}} = -\frac{4\pi^{\rm{2}} Gm}{\sqrt{{k'}_{\rm{x}}^{\rm{2}} + 
{k'}_{\rm{y}}^{\rm{2}}}}\delta\left(\omega\right)\nonumber \\ && - 
\int_{-\infty}^{k'_{\rm x}}dk_{\rm{x}} {\cal R} \left(k_{\rm{x}} , k'_{\rm{x}}
\right)\frac{4\pi^{\rm{2}}Gm}{\sqrt{k_{\rm{x}}^{\rm{2}} + {k'}_{\rm{y}}^{\rm{2}}
}}\delta\left(\omega\right)\,,
\end{eqnarray}
and the corresponding surface density distribution (cf. section 9)              
\begin{eqnarray}
&&\Sigma_{\rm{\bf{k'}}} = 2\pi m\delta\left(\omega\right) \nonumber \\ && + 
\sqrt{{k'}_{\rm{x}}^{\rm{2}} + {k'}_{\rm{y}}^{\rm{2}}} \int_{-\infty}^{k'_x}
dk_{\rm{x}} {\cal R} \left(k_{\rm{x}} , k'_{\rm{x}}\right)
\frac{2\pi m\delta\left(\omega\right)}{\sqrt{k_{\rm{x}}^{\rm{2}} + {k'}_{\rm{y}}
^{\rm{2}}}}\,.
\end{eqnarray}
Transforming this back to time and spatial coordinates leads to
\begin{eqnarray}
&&\Sigma\left(x,y,t\right) = m \delta\left(x\right)\delta\left(y\right) + 
\frac{m}{4\pi^{\rm{2}}}\int_{-\infty}^{+\infty}dk'_{\rm{x}}
\int_{-\infty}^{+\infty}dk'_{\rm{y}} \nonumber \\ &&
\int_{-\infty}^{k'_{\rm x}}dk_{\rm{x}} 
\frac{\sqrt{{k'}_{\rm{x}}^{\rm{2}} + {k'}_{\rm{y}}^{\rm{2}}}}
{\sqrt{k_{\rm{x}}^{\rm{2}} + {k'}_{\rm{y}}^{\rm{2}}}}\tilde{\cal R}
\left(k_{\rm{x}} , k'_{\rm{x}}\right)e^{i\left[k'_{\rm x} x + k'_{\rm y} y
\right]}\,.
\end{eqnarray}
The first term of equation (112) represents obviously the imposed point mass, 
whereas the second term describes a stationary polarization cloud induced by 
the point mass. Exactly the same polarization clouds have been found by 
Julian \& Toomre (1966) or Toomre \& Kalnajs (1991). Fig.~5 shows a contour
diagram of the surface density distribution of such a polarization cloud 
calculated numerically by solving the Volterra equation with an inhomogeneity 
according to equation (109) by Fredholm discretization. I find the same
surprisingly high surface density of the polarization cloud as Julian \& 
Toomre (1966). In the example shown in Fig.~5 the effective mass of the 
imposed perturber is increased by a factor of about ten due to the response of
the disk.

\begin{figure} [h]
\begin{center}
\epsfxsize=8.7cm
   \leavevmode
     \epsffile{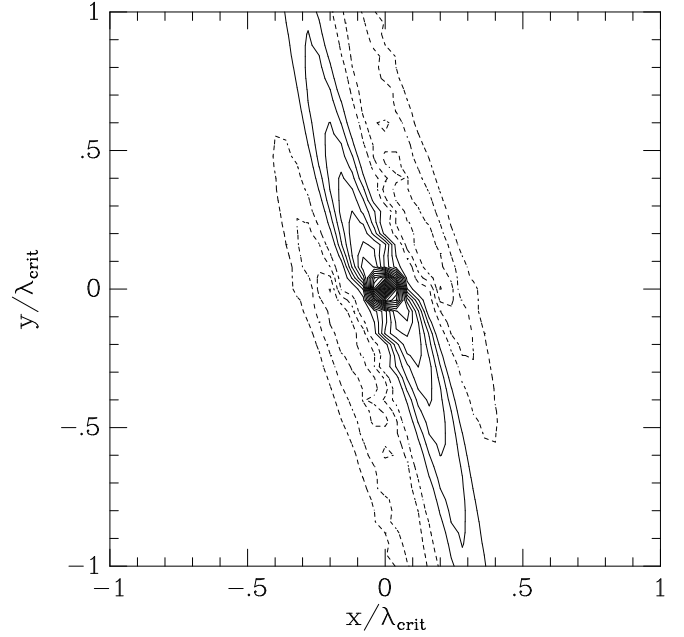}
\caption{Response of the shearing sheet to an imposed point mass illustrated by 
a contour diagram of the surface density distribution. The inner peak
represents 
the delta--function like point mass spread out according to the finite spatial 
resolution of $2\pi/k_{\rm max} = \lambda_{\rm crit}/ 8$ (FWHM). Contour levels
at multiples of 25 $m/\lambda_{\rm crit}^2$ are shown, where $m$ denotes the
imposed point mass. Negative density perturbations are drawn as dashed lines.}
\label{fig5}
   \end{center}
   \end{figure}

\subsection{Passage of a point mass}

The dynamical effect of a satellite galaxy encountering a galactic disk on an
unbound coplanar orbit can be simulated in the shearing sheet by considering a
point mass moving at constant speed $v_0$ along the y--axis. The external
potential perturbation is then given by
\begin{equation}
\Phi^{\rm{ext}} = -\frac{Gm}{\sqrt{x^{\rm{2}} + \left(y - 
v_{\rm{0}} t\right)^{\rm{2}}}}\,.
\end{equation}
Using again the formulae (3.754) and (6.671) of Gradshteyn \& Ryzhik (1965) the
potential perturbation can be Fourier transformed as
\begin{equation}
\Phi_{\rm{\bf{k'}}}^{\rm{ext}} = -\frac{2\pi Gm}{\sqrt{{k'}_{\rm{x}}^{\rm{2}} + 
{k'}_{\rm{y}}^{\rm{2}}}} e^{-ik'_{\rm y} v_0 t}\,,
\end{equation}
leading to an inhomogeneity of the Volterra equation of the form
\begin{equation}
r_{\rm{\bf{k'}}} = -\frac{4\pi^{\rm{2}} Gm}{\sqrt{{k'}_{\rm{x}}^{\rm{2}} +
{k'}_{\rm{y}}^{\rm{2}}}}\delta\left(\omega + k'_{\rm{y}} v_{\rm{0}} \right)\,.
\end{equation}
The solution of the Volterra equation has then an analogous form to equations 
(110) and (111). Transforming this back to time and spatial coordinates leads to
\begin{eqnarray}
&&\Sigma\left(x,y,t\right) = m \delta\left(x\right)\delta\left(y - v_{\rm{0}}
t\right) + \frac{m}{4\pi^{\rm{2}}}\int_{-\infty}^{+\infty}
dk'_{\rm{x}} \int_{-\infty}^{+\infty}dk'_{\rm{y}}\nonumber \\ &&
\int_{-\infty}^{k'_x}
dk_{\rm{x}} \frac{\sqrt{{k'}_{\rm{x}}^{\rm{2}} + k_{\rm{y}}\prime^{\rm{2}}}}
{\sqrt{k_{\rm{x}}^{\rm{2}} + k_{\rm{y}}^{\rm{2}}}} \tilde{\cal R}
\left(k_{\rm{x}} , k'_{\rm{x}}\right)e^{-i\left[\frac{v_0}{2A}\left(k_{\rm x}
- k'_{\rm x}\right) + k'_{\rm y} v_0 t\right]} 
\nonumber \\ && \cdot e^{i\left[k'_{\rm x} x 
- k'_{\rm y} y\right]}\,,
\end{eqnarray}
where the explicit form of the $\omega$-dependent term of the resolvent
operator $\cal R$ has been used (cf. section 7). Again the first term of 
equation (116) represents the imposed moving point mass, while the second term 
is now proportional to $e^{ik'_{\rm y}\left(y - v_0 t\right)}$, and 
represents thus a travelling polarization cloud which is stationary in the 
reference frame moving along with the point mass.

\subsection{Response to white noise}                                            

Toomre (1990) has argued that much of the recurrent spiral activity seen by him 
in his numerical sliding grid simulations of the dynamical evolution of a star
disk or by Sellwood \& Carlberg (1984) in their numerical simulations is due to
swing amplified random fluctuations induced by particle noise. In order to
illustrate this effect I have subjected also the shearing sheet to such
potential perturbations. I consider for this purpose the Fourier transform of 
the Volterra equation (68) with respect to time,
\begin{equation}
\Phi_{\rm{{\bf{k'}},t}}\left(t\right) =
\int_{-\infty}^{k'_{\rm x}}dk_{\rm{x}} \tilde{\cal K} \left(
k_{\rm{x}} ,k'_{\rm{x}}\right)\Phi_{\rm{k_{\rm x} ,k'_{\rm y},t + 
\frac{k_{\rm{x}} - k'_{\rm{x}}}{2Ak'_{\rm{y}}}}} +
 r_{\rm{{\bf{k'}},t}}\,,
\end{equation}
where use has been made of the explicit form of the $\omega$-dependent term of
the kernel function and the convolution theorem of the Fourier transform of
products of functions. This can be rewritten with the help of equation (90) 
in terms of the surface density
\begin{eqnarray}
&&\Sigma_{\rm{{\bf{k'}},t}} = \sqrt{{k'}_{\rm{x}}^{\rm{2}} + {k'}_{\rm{y}}^{\rm{
2}}}\int_{-\infty}^{k'_{\rm x}}dk_{\rm{x}} \frac{\tilde{\cal K}\left(k_{\rm{x}}
 ,k'_{\rm{x}}\right)}{\sqrt{k_{\rm{x}}^{\rm{2}} +                              
{k'}_{\rm{y}}^{\rm{2}}}}\Sigma_{\rm{k_{\rm x} ,k'_{\rm y}, t +                  
\frac{k_{\rm{x}} - k'_{\rm{x}}}{2Ak'_{\rm{y}}}}}\nonumber \\ && + 
\Sigma_{\rm{{\bf{k'}},t}}^{\rm{ext}}\,,
\end{eqnarray}
where $\Sigma_{\rm{{\bf{k'}}, t}}^{\rm{ext}}$ denotes the surface density 
associated with the randomly fluctuating potential perturbations.

In the case of white noise the autocorrelation function of the fluctuating
surface density of the sheet is given by
\begin{eqnarray}
&& \langle\Sigma\left({\bf x},t\right) \Sigma^* \left({\bf x'} ,
t \right)\rangle = \int d^{\rm{2}}k \int d^{\rm{2}}k' \langle\Sigma_{\bf{k}}
\left(t\right),\Sigma_{\bf k'}^* \left(t\right)\rangle \nonumber \\ &&
\cdot \exp i \left[\left({\bf k},{\bf x}\right) - \left({\bf k'}, {\bf x'} 
\right)\right]\,,
\end{eqnarray}
where 
\begin{equation}
\langle \Sigma_{\rm{\bf{k}}}\left(t\right),\Sigma_{\rm{\bf{k'}}}^* 
\left(t\right)\rangle = | \Sigma_{\rm{\bf{k}}}\left(t\right)|^{\rm{2}}
\delta\left(({\bf k}-{\bf k'})/k_{\rm{crit}}\right)\,,
\end{equation}
so that
\begin{eqnarray}
&& \langle\Sigma\left({\bf x},t\right) \Sigma^* \left({\bf x'},
t\right)\rangle \nonumber \\ &&
= k_{\rm{crit}}^{\rm{2}} \int d^{\rm{2}} k |\Sigma_{\rm{\bf{k}}}
\left(t\right)|^{\rm{2}} \exp -i\left({\bf k} , {\bf x}-{\bf x'}
\right)
 \nonumber \\ &&
= \left(2\pi\right)^{\rm{2}} k_{\rm{crit}}^{\rm{2}} |\Sigma_{\rm{\bf{k}}}
\left(t\right)|^{\rm{2}} \delta \left(x-x'\right) 
\delta\left(y-y'\right)\,,
\end{eqnarray}
because the amplitudes $\Sigma_{\bf{k}}$ are uniformly distributed over wave
number space. If the amplitude of the surface density fluctuations is expressed
as fraction of the background density of the unperturbed sheet, 
$\varepsilon\Sigma_{\rm{0}}$, one obtains
\begin{eqnarray}
&&\langle\Sigma\left({\bf x},t\right)\Sigma^*\left({\bf x'}
,t\right)\rangle = \varepsilon^{\rm{2}} \Sigma_{\rm{0}}^{\rm{2}} 
\nonumber\\ &&
\approx k_{\rm{crit}}^{\rm{2}} \left|\Sigma_{\rm{\bf{k}}}\left(t\right)
\right|^{\rm{2}} \frac{2\pi}{\Delta x}\frac{2\pi}{\Delta y} 
= k_{\rm{crit}}^{\rm{2}} \left|\Sigma_{\rm{\bf{k}}}\left(t\right)
\right|^{\rm{2}} \Delta k_{\rm{x}} \Delta k_{\rm{y}}\,,
\end{eqnarray}
where the delta functions of equation (121) have been approximated according to 
the finite resolution of the numerical solution of the Volterra equation (118).
The inhomogeneity of the integral equation is then given by
\begin{equation}
\Sigma_{\rm{\bf{k}}}^{\rm{ext}} = \left[ \frac{\varepsilon^{\rm{2}} 
\Sigma_{\rm{0}}^{\rm{2}}}{k_{\rm{crit}}^{\rm{2}} \Delta k_{\rm{x}} 
\Delta k_{\rm{y}}}\right]^\frac{1}{2} \cdot \cos\left(\phi\right)\,,
\end{equation}
defined here as a real quantity, and where $\phi$ denotes a random angle
distributed uniformly over the range $\left[0,2\pi\right]$. Finally it is 
advantageous to write the integral equation (118) and expression (123) in       
dimensionless form by dividing them by the background density, which in wave 
number space is given by $\Sigma_{\rm{0}} \delta\left(\bf{k}\right)
\approx \Sigma_{\rm{0}} / \left(\Delta k_{\rm{x}} \Delta k_{\rm{y}} \right)$.
Fig.~6 shows a typical solution of equation (118). The parameters of the disk 
model are adopted again as $A/\Omega_{\rm{0}} = 0.5$ and $Q = 1.4$. The
time steps at which solutions of equation (118) can be calculated are given by
multiples of
\begin{equation}
\Delta t = \frac{\Delta k_{\rm{x}}}{2Ak'_{\rm{y}}}\,,
\end{equation}
depending on the circumferential wave number $k'_{\rm y}$. Thus solutions are 
obtained at 
different time intervals. However, within less than an orbital period a
quasi--stationary equilibrium is reached, which is shown in the second panel 
of Fig.~6. Each spike represents the amplitude
$\sqrt{|\Sigma_{\rm{\bf k}} |^{\rm{2}}}$ of the superposition of the randomly
incited swing-amplified density waves, all travelling from left to right at a 
velocity of $\dot{k}_{\rm{x,eff}} = 2Ak_{\rm{y}}$. The contour plot in Fig.~6
shows clearly the peak of amplitudes around
$\left(k_{\rm{x}} ,k_{\rm{y}}\right) \approx $ (1.7 , 0.5)
$k_{\rm{crit}}$. This corresponds exactly to the amplification 
factors shown in Fig.~1.
\begin{figure} [h]
\begin{center}
\epsfxsize=8.7cm
   \leavevmode
     \epsffile{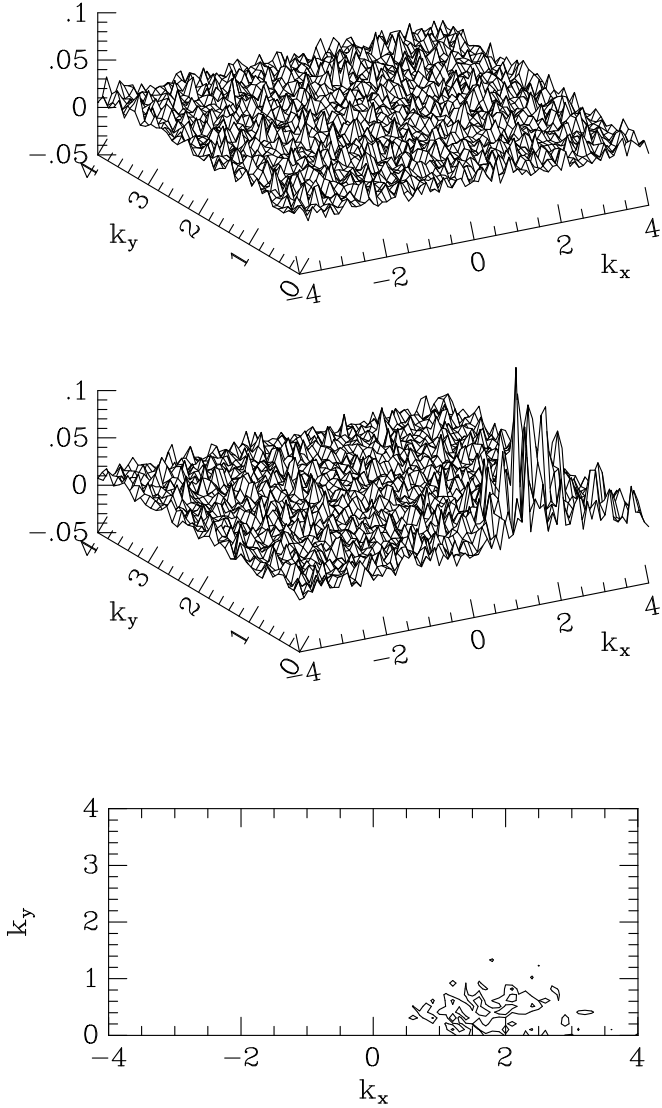}
\caption{Response of the swing amplifier to white noise input. The amplitude of
the surface density, $\sqrt{|\Sigma_{\rm{\bf k}}|^2}$, is plotted versus wave
numbers
$k_{\rm x}$ and $k_{\rm y}$. The upper panel shows a snapshot of the ever
present, externally imposed white noise fluctuations. The middle panel shows the
quasi--stationary equilibrium reached after the perturbations of the shearing 
sheet have fully developed. The lower panel shows a contour plot of the
quasi--stationary equilibrium distribution. The parameters of the disk model are
chosen as $A/\Omega_0$=0.5 and $Q$=1.4.}  
\label{fig6}
   \end{center}
   \end{figure}
It was already noted in the numerical simulations by
Sellwood \& Carlberg (1984) that the constant spiral activity leads to dynamical
heating of the disk, because the stars are randomly accelerated by the shearing
density waves. This implies a rapid increase of the velocity dispersion of the
stars and thus of the stability parameter $Q$ (cf.~equation 95), which will
eventually suppress any spiral activity, if the disk is not dynamically cooled.
Disk heating by swing amplified density waves in the shearing sheet will be
discussed in detail in a further paper of this series (Fuchs 2001, in 
preparation).

\section{Summary and conclusion}
The objective of this paper is to present a new description of the dynamics of
a shearing sheet made of stars. It is shown again that disturbances of
the disk evolve always into swing
amplified density waves, i.e. spiral--arm like, shearing density waves, which
grow and decay while their wave crests swing by from leading to trailing
orientation. Specific examples are discussed how such swing amplification events
are incited. Only in response to imposed massive perturbers the shearing sheet
develops stationary perturbations.

\acknowledgements{I thank A. Toomre and R. Wielen for their advice over the
years. Helpful discussions with E. Athanassoula, N.W. Evans, J. Sellwood, and
M. Tagger are gratefully acknowledged.}

{}

\section*{Appendix}

Comparison of equations (89) and (87) shows that the amplitude function in
equation (89) can be calculated directly from the solution of the Volterra 
integral equation (68) with the inhomogeneity (86), if $k'_{\rm x}$ takes the
role of $k_{\rm x}^{\rm in} + 2 A k_{\rm y}^{\rm in} t$, ($ k'_{\rm x} \geq 
{k'}_{\rm x}^{\rm in}$), and $\cal K$ and $\cal L$ are replaced by
$\tilde{\cal K}$ and $\tilde{\cal L}$, ($\omega = 0$),
\begin{equation}
\Phi_{\rm k'_{\rm x}, k'_{\rm y}} = \int_{\rm k_{\rm x}^{\rm in}}^{\rm k'_{\rm
x}} dk_{\rm x} \tilde{\cal K}(k_{\rm x}, k'_{\rm x})
\Phi_{\rm k_{\rm x}, k'_{\rm y}} + \tilde{\cal L}(k_{\rm x}^{\rm in},
 k'_{\rm x})\,.
\end{equation}
Equation (125) is discretized as ($ k'_{\rm x} = k'_{\rm j}$)
\begin{eqnarray}
&&\int_{\rm k_{\rm i}}^{\rm k_{\rm i}+1} dk_{\rm x}\tilde{\cal K}
(k_{\rm x}, k'_{\rm x})\Phi( k_{\rm x}, k'_{\rm y}) =\nonumber \\ &&
\frac{1}{2}\Big[ \tilde{\cal K}(k_{\rm i}+1, k'_{\rm j})
\Phi( k_{\rm i}+1, k'_{\rm y})+\tilde{\cal K}(k_{{\rm i}}, k'_{\rm j})
\Phi( k_{{\rm i}}, k'_{\rm y}) \Big] \Delta k
\end{eqnarray}
leading to 
\begin{eqnarray}
&&\Phi( k'_{{\rm j}}, k'_{\rm y}) =\sum_{k_{\rm i}=-\infty}^{k'_{\rm j}-1}
\frac{\Delta k}{2}\Big[ \tilde{\cal K}(k_{\rm i}+1, k'_{\rm j})
\Phi( k_{\rm i}+1, k'_{\rm y})
\nonumber \\ && +\tilde{\cal K}(k_{{\rm i}}, k'_{\rm j})
\Phi( k_{{\rm i}}, k'_{\rm y}) \Big] +\tilde{\cal L}(k_{\rm x}^{\rm in},
k'_{{\rm j}})\,.
\end{eqnarray}
According to the definition of the kernel $\tilde{\cal K}(k'_{\rm j}, k'_{\rm j}
) = 0$, so that the set of algebraic equations (127) can be resolved 
recursively:
\begin{eqnarray}
&& k'_{\rm j} < k_{\rm x}^{\rm in}:\,\Phi( k'_{\rm j}, k'_{\rm y}) = 0,
\nonumber
\\ && k'_{\rm j} = k_{\rm x}^{\rm in}:\,\Phi( k'_{\rm j}, k'_{\rm y}) = 
\tilde{\cal L}(k_{\rm x}^{\rm in}, k_{\rm x}^{\rm in} ), \nonumber \\ &&
k'_{\rm j} = k_{\rm x}^{\rm in}+1:\,\Phi( k'_{\rm j}, k'_{\rm y}) = 
\Big[ \tilde{\cal K}(k_{\rm x}^{\rm in}-1, k_{\rm x}^{\rm in}
+1)\Phi( k_{\rm x}^{\rm in}-1, k'_{\rm y})\nonumber \\&&
+\tilde{\cal K}(k_{\rm x}^{\rm in}, k_{\rm x}^{\rm in}
+1)\Phi( k_{\rm x}^{\rm in}, k'_{\rm y})+\tilde{\cal K}(k_{\rm x}^{\rm in}, 
k_{\rm x}^{\rm in}+1)\Phi( k_{\rm x}^{\rm in}, k'_{\rm y})\nonumber \\&&
+\tilde{\cal K}(k_{\rm x}^{\rm in}+1, k_{\rm }^{\rm in}+1)
\Phi( k_{\rm x}^{\rm in}+1, k'_{\rm y})\Big] \frac{\Delta k}{2} \nonumber \\ &&
+\tilde{\cal L}
(k_{\rm x}^{\rm in}, k_{\rm x}^{\rm in}+1)\nonumber \\ &&
= \tilde{\cal L}(k_{\rm x}^{\rm in}, k_{\rm x}^{\rm in}+1)+\tilde{\cal K}
(k_{\rm x}^{\rm in}, k_{\rm }^{\rm in}+1)\Phi( k_{\rm x}^{\rm in}, k'_{\rm y})
\Delta k, \nonumber \\ &&
k'_{\rm j} > k_{\rm x}^{\rm in}:\,
\Phi( k'_{\rm j}, k'_{\rm y}) =\tilde{\cal L}(k_{\rm x}^{\rm in},  k'_{\rm j})
\nonumber \\ && +\sum_{k_{\rm i} = k_{\rm x}^{\rm in}}^{k'_{\rm j}-1}
\tilde{\cal K}(k_{\rm i}, k'_{\rm j})\Phi( k_{\rm i}, k'_{\rm y})\Delta k \,.
\end{eqnarray}
In the numerical realization of the scheme I have adopted a stepwidth of 
$\Delta k$ = 0.1 $k_{\rm crit}$.

\end{document}